\documentclass[12pt]{article}

\title{}
\author{}
\date{}
\usepackage{amsmath}
\usepackage{mathtools} 
\usepackage{amssymb}
\usepackage{bm}
\usepackage{siunitx}
\usepackage{gensymb}
\usepackage[margin=1in]{geometry}
\usepackage[document]{ragged2e}
\usepackage{caption} 
\usepackage{graphicx}
\usepackage{grffile}
\usepackage{subfig}
\usepackage[T1]{fontenc}
\usepackage{mathptmx}
\usepackage{setspace}
\usepackage{hanging}
\usepackage{xcolor}
\usepackage[normalem]{ulem}
\usepackage{multirow}
\usepackage{tikz}
\usetikzlibrary{shapes.geometric}
\usetikzlibrary{arrows.meta,arrows}

\title{Partial identification for discrete data with nonignorable missing outcomes.}
\author{Daniel Daly-Grafstein and Paul Gustafson}
\date{ \vspace{-1.25cm} Department of Statistics, University of British Columbia \\
\vspace{0.25cm}
\today}

\begin{document}
\maketitle

\textbf{Abstract} \, 

Nonignorable missing outcomes are common in real world datasets and often require strong parametric assumptions to achieve identification. These assumptions can be implausible or untestable, and so we may forgo them in favour of partially identified models that narrow the set of $\textit{a priori}$ possible values to an identification region. Here we propose a new nonparametric Bayes method that allows for the incorporation of multiple clinically relevant restrictions of the parameter space simultaneously. We focus on two common restrictions, instrumental variables and the direction of missing data bias, and investigate how these restrictions narrow the identification region for parameters of interest. Additionally, we propose a rejection sampling algorithm that allows us to quantify the evidence for these assumptions in the data. We compare our method to a standard Heckman selection model in both simulation studies and in an applied problem examining the effectiveness of cash-transfers for people experiencing homelessness.

\section{Introduction}

Missing outcomes are a common feature of epidemiological data. Missingness is commonly categorized as missing at random (MAR) or nonignorable missingness, where in the latter case the probability of missingness depends on the missing values themselves (Rubin 1976). When data have nonignorable missingness, the missing data mechanism must be incorporated into the model when performing inference. 

A fundamental issue when modelling the missing data mechanism is that we do not observe any data to estimate its distribution. Therefore, parameters of interest from data with nonignorable missing outcomes usually are not identified without incorporating strong parametric assumptions into the model. Several classes of models have been developed that can lead to identification without resorting to a missing at random assumption, including sample selection models (Heckman 1979, Van de Ven and Van Praag 1981), pattern mixture models (Little 1993), and joint shared parameter models (Hogan and Daniels 2008). All of these approaches can lead to identification given certain parametric assumptions, however it is difficult to evaluate these assumptions, and estimates can be poor when parametric assumptions are violated (McGovern et al. 2016).  

If we are uncertain about parametric identification restrictions for our missing data we can instead use a nonidentified model where parameter estimates converge to an identification set, rather than a single value, in the limit. In a Bayesian context, we can separate our posterior distribution into parameters for the observed data, and parameters for missing data conditional on the observed data. Recently, nonparametric Bayesian approaches have been proposed that specify flexible nonparametric models for the observed data and allow for more intuitive, clinically relevant assumptions to be placed on the missing data distribution (Linero and Daniels 2015, Linero and Daniels 2018). 

Assumptions on the missing data mechanism in this manner may not lead to consistent estimation of all parameters, but can lead to partial identification where there is a  narrowing of the identification set relative to the set of $\textit{a priori}$ possible values (Gustafson 2015). One common assumption is the presence of an instrumental variable. An instrumental variable in the context of nonignorable missing outcomes is defined similarly to the usual causal inference setting (Marden et al. 2019), namely we assume the instrument is only related to the outcome through the treatment variable, and the instrument is associated with the missingness mechanism (Figure 1). Instrumental variable assumptions have been shown to shrink identification regions in problems with nonignorable missingness (Manski 2005, Mattei et al. 2014, Marden et al. 2019). 

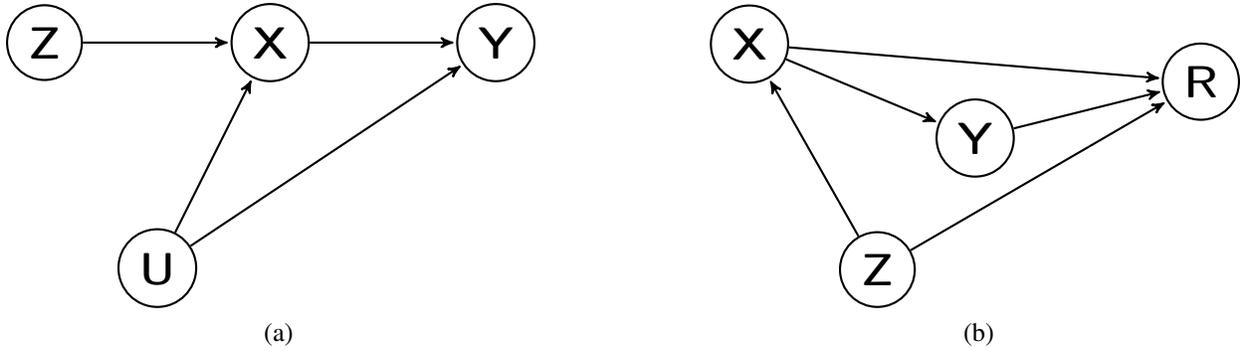
\begin{figure}
  \centering
  \subfloat[]{%
  \begin{tikzpicture}[->,>=stealth',shorten >=1pt,auto,node distance=3cm,
                    thick,main node/.style={circle,draw,font=\sffamily\Large\bfseries}]

  \node[main node] (Z) {Z};
  \node[main node] (X) [right of=Z] {X};
  \node[main node] (U) [below of=X, xshift=-1.5cm] {U};
  \node[main node] (Y) [right of=X] {Y};

  \path[every node/.style={font=\sffamily\small}]
    (Z) edge node {} (X)
    (U) edge node {} (X)
    (X) edge node {} (Y)
    (U) edge node {} (Y);
\end{tikzpicture}
    \label{fig:subfig1}
  }
  \hfill
  \subfloat[]{%
  \begin{tikzpicture}[->,>=stealth',shorten >=1pt,auto,node distance=3cm,
                    thick,main node/.style={circle,draw,font=\sffamily\Large\bfseries}]

  \node[main node] (Y) {Y};
  \node[main node] (X) [left of =Y, yshift=1.25cm] {X};
  \node[main node] (Z) [below of=X, xshift=1.7cm] {Z};
  \node[main node] (R) [right of=Y, yshift=0.75cm] {R};

  \path[every node/.style={font=\sffamily\small}]
    (Z) edge node {} (X)
    (X) edge node {} (Y)
    (Y) edge node {} (R)
    (Z) edge node {} (R)
    (X) edge node {} (R);
\end{tikzpicture}
    \label{fig:subfig2}
  }
  \caption{Directed acyclic graphs depicting instrumental variables in the context of unmeasured confounding (a) and non-ignorable missing outcome data (b). Here $X$ denotes a treatment, $Y$ a potentially missing outcome, $Z$ an instrumental variable, $U$ unmeasured confounding, and $R$ a missingness indicator.}
  \label{fig:overall}
\end{figure}

In this paper we propose a new nonparametric Bayes method for data with nonignorable missing outcomes. Our method extends the work of Mattei et al. (2014) by allowing the specification multiple prior assumptions simultaneously, as well as the ability to partially relax instrumental variable or other $\textit{a priori}$ missing data assumptions as needed. We focus on settings with nonignorable missing binary outcomes and binary treatment variables, which are common in randomized control trials (RCT) where nonignorable missingness occurs after treatment assignment. Additionally, a rejection sampling technique is proposed that allows us to quantify the evidence present in the data for a given set of restricting assumptions. We compare our method using a variety of prior restrictions to a sample selection model that requires parametric assumptions in two simulation studies. We then apply our method to data from a RCT that examines the impact of unconditional cash transfers on people experiencing homelessness.

\subsection{Notation and problem setting}

We motivate our method using a simple setting with discrete data. Let $Y$ denote a binary outcome, $X$ a single binary treatment variable, and $Z$ a single binary variable that may be an instrument. We are interested in the effect of $X$ on $Y$ defined as $\Psi = P(Y|X=1) - P(Y|X=0)$. In this setting we observe all $X$ and $Z$, but only some proportion of $Y$. Let $R$ denote a binary indicator variable, with $R_i=1$ indicating $Y_i$ is observed. If we assume $R \perp Y|(X,Z)$, then the data are MAR, however in our case this does not hold and the data contain nonignorable missingness. In cases where $Z$ is truly an instrument, we have $Y \perp Z|X$ while having $R \not\perp Z|X$. When $Z$ is not an instrument, we will relax the assumption that $Z$ only influences $Y$ through $X$, namely $Y \not\perp Z|X$.

Since we have nonignorable missingness we will consider the joint distribution $P(Y, R|X,Z)$. We let $p_{x,z,r,y} = P(Y=y, R=r|X=x, Z=z)$ and $p_{x,z} = \{p_{x,z,1,0}, p_{x,z,1,1}, p_{x,z,0,0}, p_{x,z,0,1}\}$. In cases where we are referring to some marginalization of this distribution, for example the probability of missingness $P(R=r|X=x, Z=z)$, we denote this as $p_{x,z,r,\boldsymbol{\cdot}}$. Finally, we denote the true value of parameters under which data are generated with the dagger symbol $\boldsymbol{\alpha}^\dagger$.

\subsection{Sample selection model}

Throughout the paper we will compare our nonparametric method to a classical sample selection model that parametrically identifies $\Psi$ in the presence of nonignorable missingness (Heckman 1979,  Van de Ven and Van Praag 1981). In our setting the selection model assumes a bivariate probit distribution for $Y$ and $R$ with a latent bivariate normal distribution

\begin{align*}
Y_i^* =& \beta_{0} + \beta_{1}X_i + \epsilon_i  \\[5pt]
R_i^* =& \gamma_0 + \gamma_1 X_i + \gamma_2 Z_i + \nu_i\\[8pt]
\begin{pmatrix}\epsilon\\
\nu
\end{pmatrix} &\sim  N
\begin{bmatrix}
\begin{pmatrix}
0\\
0
\end{pmatrix},
\begin{pmatrix}
1 & \rho \\
\rho & 1 
\end{pmatrix}
\end{bmatrix}
\end{align*}

where $Y_i = 1$ if $Y_i^* > 0$ and $R_i = 1$ if $R_i^* > 0$. The correlation coefficient $\rho$ governs the dependence between missingness $R$ and outcome $Y$, with $\rho=0$ corresponding to the MAR assumption. While in theory the sample selection model consistently estimates $\Psi$, in finite-sample settings it has been shown that $Z_i$ is needed as an instrument to avoid collinearity issues (Puhani 2000). This results in marginals for $R$ and $Y$ of 

\begin{align*}
P(Y_i = 1|X_i, Z_i) =& \Phi(\beta_{0} + \beta_{1} X_i) \\
P(R_i = 1|X_i, Z_i) =& \Phi(\gamma_0 + \gamma_1X_i + \gamma_2{Z_i}),
\end{align*}

where $\Phi$ is the standard normal cumulative distribution.

\section{Nonparametric modelling}

The starting point of our method is a fully saturated approach where we make no parametric assumptions about $X$, $Z$, or $Y$, and no instrumental variable assumptions on $Z$. We construct a saturated model for $(Y, R|X, Z)$ where we separately parameterize each combination of the discrete variables $X, Z$. For each $p_{x,z}$ we assume independent Dirichlet priors such that $p_{x,z} \sim Dir(\alpha_1, \alpha_2, \alpha_3, \alpha_4$) for all $X=x, Z=z$. 

If we observe multinomial data for $Y$ and $R$ we can update $p_{x,z}$ in a conjugate fashion. However, with missing data we do not observe $Y$ when $R=0$, so we have no direct information about $p_{x,z,0,0}$ or $p_{x,z,0,1}$, only $p_{x,z,0, \boldsymbol{\cdot}}$. It will be useful for us to separate parameters into those parameterizing the observed data distribution and those parameterizing the missing data distribution to allow for more interpretable missing data assumptions. We can reparametrize our prior on $p_{x,z}$ as:

\begin{align*}
g_1(p_{x,z}) &= p_{x,z,1,0} \coloneqq q_{x,z,1,0}\\
g_2(p_{x,z}) &= p_{x,z,1,1} \coloneqq q_{x,z,1,1}\\
g_3(p_{x,z}) &= p_{x,z,0,0} + p_{x,z,0,1} \coloneqq q_{x,z,0,\boldsymbol{\cdot}} \\
g_4(p_{x,z}) &= \frac{p_{x,z,0,1}}{p_{x,z,0,0} + p_{x,z,0,1}} \coloneqq \omega_{x,z}
\end{align*}

giving us $[q_{x,z,1,0}, q_{x,z,1,1}, q_{x,z,0,\boldsymbol{\cdot}}] \sim Dir(\alpha_1, \alpha_2, \alpha_3 + \alpha_4)$ and $[\omega_{x,z}] \sim Beta(\alpha_3, \alpha_4)$ independently, where $q_{x,z,0,\boldsymbol{\cdot}} = P(Y=1,R=0|X=x,Z=z) + P(Y=0,R=0|X=x,Z=z)$. If we begin with a $Dir(1,1,1,1)$ prior, then the prior distribution on $\omega_{x,z}$ becomes a $Unif(0,1)$. We will make this assumption throughout the paper. 

Our observed data depends on $(q_{x,z,1,0}, q_{x,z,1,1}, q_{x,z,0,\boldsymbol{\cdot}})$, with $q_{x,z,0,\boldsymbol{\cdot}}$ denoting the total proportion of missing $Y$ values. The $\omega_{x,z}$ parameters are not directly estimable and represent the proportion of missing data where $Y=1$. This results in a joint prior for all $q_{x,z}, \omega_{x,z}$ of 

\begin{align}
f(q_{0,0},q_{0,1},q_{1,0},q_{1,1},\omega_{0,0},\omega_{0,1},\omega_{1,0},\omega_{1,1}) = &f(q_{0,0} ; \bm{\alpha}_{0,0}) f(q_{0,1} ; \bm{\alpha}_{0,1})f(q_{1,0} ; \bm{\alpha}_{1,0})f(q_{1,1} ; \bm{\alpha}_{1,1}) \nonumber\\
&I_{(0,1)}(\omega_{0,0})I_{(0,1)}(\omega_{0,1})I_{(0,1)}(\omega_{1,0})I_{(0,1)}(\omega_{1,1}),
\end{align}

where $q_{x,z} = (q_{x,z,1,0}, q_{x,z,1,1}, q_{x,z,0,\boldsymbol{\cdot}})$, $\bm{\alpha}_{x,z} = (1, 1, 2)$, and $f(q_{x,z} ; \bm{\alpha}_{x,z})$ is the probability density function of a Dirichlet distribution. We will refer to this as the "Sat" model. 

\subsection{Instrumental variable assumptions and nonparametric bounds}

Without assumptions on the missing data parameters $\omega_{x,z}$ the nonparametric bounds of our parameter of interest $\Psi$ may be too wide to be informative when the proportion of missingness is high. We can assume $\textit{a priori}$ restrictions on the $\omega_{x,z}$'s that narrow the bounds of $\Psi$. For example, assuming $Z$ is a true instrumental variable implies $P(Y=1|X,Z=1) = P(Y=1|X, Z=0)$. We can specify this restriction in (1) by restricting the $(Y, Z)$ odds ratio to 1 when $X=0$ and $X=1$ as

\begin{align}
f(q_{0,0},q_{0,1},q_{1,0},q_{1,1},\omega_{0,0},\omega_{0,1},\omega_{1,0},\omega_{1,1}) \propto &f(q_{0,0} ; \bm{\alpha}_{0,0}) f(q_{0,1} ; \bm{\alpha}_{0,1})f(q_{1,0} ; \bm{\alpha}_{1,0})f(q_{1,1} ; \bm{\alpha}_{1,1}) \nonumber \\
&I_{(0,1)}(\omega_{0,0})I_{(0,1)}(\omega_{0,1})I_{(0,1)}(\omega_{1,0})I_{(0,1)}(\omega_{1,1}) \\ 
&I\{OR(Y,Z|X=1) = 1\} I\{OR(Y,Z|X=0) = 1\},  \nonumber
\end{align}

where, for example, $I\{OR(Y,Z|X=0) = 1\}$ can be written as

\begin{equation}
 (q_{0,0,0,\boldsymbol{\cdot}}\omega_{0,0} + q_{0,0,1,1}) -  (q_{0,1,0,\boldsymbol{\cdot}}\omega_{0,1} + q_{0,1,1,1}) = 0.
\end{equation}

To see how this restricts our missing data parameters $\omega_{x,z}$, without loss of generality consider the joint prior (2) for only parameters in the $X=0$ stratum, as parameters for $X=1$ and $X=0$ are independent. If we take transformations ${\omega}_{0,0}^* = (q_{0,0,0,\boldsymbol{\cdot}}\omega_{0,0} + q_{0,0,1,1}) -  (q_{0,1,0,\boldsymbol{\cdot}}\omega_{0,1} + q_{0,1,1,1}) $ and ${\omega}_{0,1}^* = q_{0,1,0,\boldsymbol{\cdot}}\omega_{0,1} + q_{0,1,1,1}$, we can see the instrumental variable assumption is equivalent to assuming $\omega_{0,0}^* = 0$. Thus the joint prior density with this assumption is

\begin{align}
f(q_{0,0},q_{0,1},\omega_{0,1}^*, \omega_{0,0}^* = 0) &= f(q_{0,0},q_{0,1},\omega_{0,1}^*| \omega_{0,0}^* = 0)f(\omega_{0,0}^* = 0) \propto \nonumber \\
f_{q_{0,0}}f_{q_{0,1}}\frac{1}{q_{0,0,0,\boldsymbol{\cdot}}}\frac{1}{q_{0,1,0,\boldsymbol{\cdot}}}&I_{\left(max \left(0,\frac{q_{0,0,1,1} - q_{0,1,1,1}}{q_{0,1,0,\boldsymbol{\cdot}}} \right),min \left(1,\frac{q_{0,0,1,1} - q_{0,1,1,1} + q_{0,0,0,\boldsymbol{\cdot}}}{q_{0,1,0,\boldsymbol{\cdot}}}\right) \right)}(\omega_{0,1})\nonumber \\ 
&\delta_{\omega_{0,0}}\left( \frac{q_{0,1,1,1} - q_{0,0,1,1} + q_{0,1,0,\boldsymbol{\cdot}}\omega_{0,1}}{q_{0,0,0,\boldsymbol{\cdot}}} \right),
\end{align}

and we will refer to (4) as the "Sat-OR-1" model. Note this is not the only way to parameterize the instrumental variable assumption. For example, we could instead set $\omega_{0,0}^* = (q_{0,0,0,\boldsymbol{\cdot}}\omega_{0,0} + q_{0,0,1,1})/(q_{0,1,0,\boldsymbol{\cdot}}\omega_{0,1} + q_{0,1,1,1})$ and condition on $\omega_{0,0}^* = 1$. While equally valid the latter results in a different, more complex, prior density due to the Borel–Kolmogorov paradox (Rescorla 2015). 

Comparing the Sat and Sat-OR-1 priors we can see how an instrumental variable assumption on $Z$ narrows the bounds for $\omega_{x,z}$. The conditional priors on these parameters go from $Unif(0,1)$ distributions to 

\begin{align*}
[\omega_{0,1}|{q_{x,z}} ] &\sim Unif \left (max \left (0, \frac{q_{0,0,1,1} - q_{0,1,1,1}}{q_{0,1,0,\boldsymbol{\cdot}}} \right),
min \left (1, \frac{q_{0,0,1,1} - q_{0,1,1,1} + q_{0,0,0,\boldsymbol{\cdot}}}{q_{0,1,0,\boldsymbol{\cdot}}}\right ) \right ) \\
[\omega_{1,1}|{q_{x,z}} ] &\sim Unif \left (max \left (0, \frac{q_{1,0,1,1} - q_{1,1,1,1}}{q_{1,1,0,\boldsymbol{\cdot}}} \right),
min \left (1, \frac{q_{1,0,1,1} - q_{1,1,1,1} + q_{1,0,0,\boldsymbol{\cdot}}}{q_{1,1,0,\boldsymbol{\cdot}}}\right ) \right ) \\
[\omega_{0,0}|{q_{x,z}} ] &\sim \delta_{\omega_{0,0}}\left( \frac{q_{0,1,1,1} - q_{0,0,1,1} + q_{0,1,0,\boldsymbol{\cdot}}\omega_{0,1}}{q_{0,0,0,\boldsymbol{\cdot}}} \right) \\
[\omega_{1,0}|{q_{x,z}} ] &\sim \delta_{\omega_{1,0}}\left( \frac{q_{1,1,1,1} - q_{1,0,1,1} + q_{1,1,0,\boldsymbol{\cdot}}\omega_{1,1}}{q_{1,0,0,\boldsymbol{\cdot}}} \right).
\end{align*}

Our parameter of interest $\Psi$ can be written as

\begin{equation}
\begin{aligned}
\Psi = &\left( (q_{1,1,1,1} + q_{1,1,0,\boldsymbol{\cdot}}\omega_{1,1})q_z +  
         (q_{1,0,1,1} + q_{1,0,0,\boldsymbol{\cdot}}\omega_{1,0})(1-q_z) \right) - \\
         & \left( (q_{0,1,1,1} + q_{0,1,0,\boldsymbol{\cdot}}\omega_{0,1})q_z +  
         (q_{0,0,1,1} + q_{0,0,0,\boldsymbol{\cdot}}\omega_{0,0})(1-q_z) \right),
\end{aligned}
\end{equation}

where $q_z = P(Z=1)$. The upper nonparametric bound on $\Psi$ can be found by maximizing $\omega_{1,1}, \omega_{1,0}$ and minimizing $\omega_{0,1}, \omega_{0,0}$ given our observed data parameters $q_{x,z}$, and minimized when the reverse is true. Therefore, restrictions on the intervals of $\omega_{x,z}$ by our instrumental variable assumption will narrow the nonparametric bounds on $\Psi$.

In the above case we assumed $Z$ is a true instrument, however we may believe $\textit{a priori}$ that $Z$ is an imperfect instrument with some residual dependence with $Y$ given $X$. Nonparametric bounds for treatment effects with imperfect instruments have been developed in the latent unobserved confounding context (Ban and Kedagni 2022), and we can do something similar here with missing data. To define our imperfect instrument we can restrict $OR(Y,Z|X)$ to be within some threshold $(t_l, t_h)$ around 1 rather than equal to 1, or we can induce a probability distribution over $OR(Y,Z|X=1)$ and $OR(Y,Z|X=0)$ concentrated at 0. For example if we multiply the prior in (1) by normal density functions over $log(OR(Y,Z|X))$, then the standard deviation of these normals $\sigma$ determines how close we think $Z$ to a true instrument. We refer to these models as "Sat-OR-thresh" and "Sat-OR-normal", respectively. See Section A.1 in the Appendix for further details.  

\subsection{Assumptions on the direction of missing data bias}

Initially the reparameterization of our saturated model resulted in parameters $q_{x,z}$ and $\omega_{x,z}$ cleanly separated into dependent and independent of the observed data, respectively. However the introduction of an instrumental variable assumption of $Z$ induces an $\textit{a priori}$ dependence between these variables as seen in the joint priors in the previous section. While this assumption has the potential to narrow nonparametric bounds on $\Psi$, the magnitude of this narrowing depends on our observed data.

\begin{table}[!t]
\caption{The true probabilities $q_{x,z}^\dagger$ in our two examples. Note starred values are missing, and we can only estimate the sum of columns 3 and 4 for each combination of $X$ and $Z$.}
\centering
\renewcommand{\arraystretch}{1.5}
\begin{tabular}{c|cccc}
  \hline
  Example 1  & (Y=0, R=1) &  (Y=1, R=1) &   (Y=0, R=0) &  (Y=1, R=0) \\
  \hline
  $P(Y, R|X=0, Z=0)$ & 0.47 & 0.06 & 0.11* & 0.37* \\
  \hline
  $P(Y, R|X=0, Z=1)$ & 0.54 & 0.29 & 0.03* & 0.13* \\
  \hline
  $P(Y, R|X=1, Z=0)$ & 0.29 & 0.28 & 0.14* & 0.29* \\
  \hline
  $P(Y, R|X=1, Z=1)$ & 0.38 & 0.49 & 0.04* & 0.08* \\
 & & & & \\
\hline
  \hline
  Example 2  & (Y=0, R=1) &  (Y=1, R=1) &   (Y=0, R=0) &  (Y=1, R=0) \\
  \hline
  $P(Y, R|X=0, Z=0)$ & 0.49 & 0.14 & 0.11* & 0.25* \\
  \hline
  $P(Y, R|X=0, Z=1)$ & 0.50 & 0.23 & 0.11* & 0.17* \\
  \hline
  $P(Y, R|X=1, Z=0)$ & 0.31 & 0.22 & 0.05* & 0.41* \\
  \hline
  $P(Y, R|X=1, Z=1)$ & 0.27 & 0.24 & 0.10* & 0.39* \\

  \hline
\end{tabular}
\setlength{\abovecaptionskip}{10pt}
\end{table}

Here we give two examples, one where an instrumental variable assumption significantly narrows the bounds of $\Psi$, and one where it does not. In both examples we consider inference in the limiting posterior distribution where we have infinite data and the marginal posteriors of our parameters $q_{x,z}$ are point masses at $q_{x,z}^\dagger$. Consider our first example with probabilities shown in Table 1. In this example we have a combination of large observed differences in $P(Y=1|R=1, X=x, Z=z)$ and low levels of missingness in at least one of the two cells when $X=0$ and $X=1$. If we make the assumption that $Z$ is an instrument (i.e. $OR(Y,Z|X) = 1$), this forces our $\omega$ parameters to be in restricted ranges relative to the case where we make no assumptions. Figure 2 (a) and (b) show the posteriors for $P(Y=1|X=1)$ and $log(OR(Y,Z|X=1))$ implied by the joint priors for the Sat and Sat-OR-1 models, as well as the posterior for $\Psi$. Restricting the odds ratio to 1 narrows the nonparametric bounds on $\Psi$ and sharpens inference in this case. 

\begin{figure}[!t]
    \centering
    \subfloat[]{\includegraphics[width=0.5\textwidth]{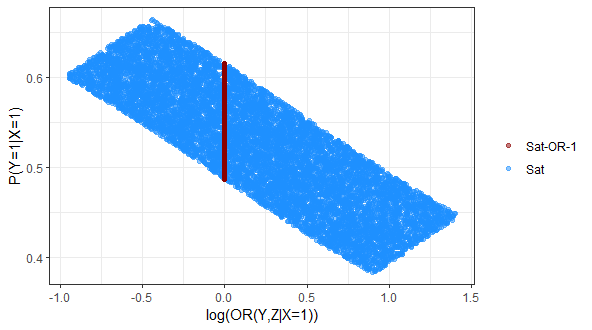}}\hfill
    \subfloat[]{\includegraphics[width=0.5\textwidth]{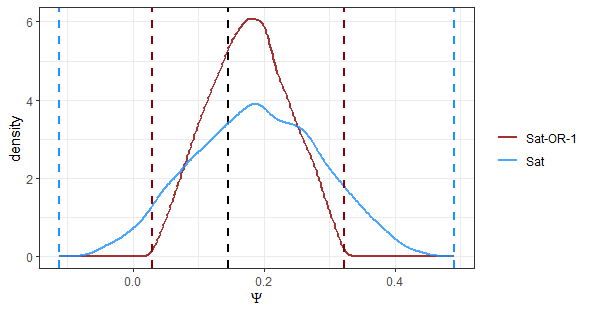}} \\
    \subfloat[]{\includegraphics[width=0.5\textwidth]{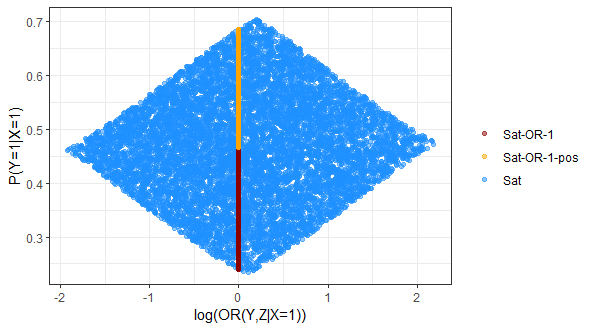}}\hfill
    \subfloat[]{\includegraphics[width=0.5\textwidth]{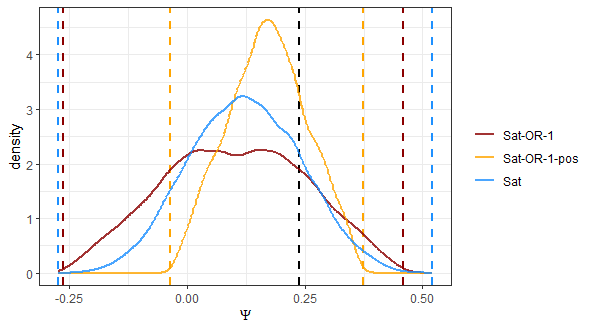}}
    \caption{Figure (a) compares the posterior distributions of $P(Y=1|X=1)$ and $log(OR(Y,Z|X=1))$, and (b) shows the posterior for $\Psi$ in example 1. Dots in (a) denote Monte Carlo draws for the posteriors implied by the Sat and Sat-OR-1 models, and dashed lines in (b) denote the bounds of the limiting posterior distribution for $\Psi$ under each assumption, with the black dashed line denoting the true value. Figures (c) and (d) repeat this for example 2, with yellow denoting the Sat-OR-1-pos model. See Section 3 for details on posterior computation.}
    \label{fig:combined}
\end{figure}

Now consider the second example shown in Table 1. Here we do not have large observed differences in $P(Y=1|R=1, X=x, Z=z)$ and all cells have moderate to high levels of missingness. The result in this case is the instrumental assumption not restricting our $\omega$ parameters much relative to the Sat model (Figure 2c, 2d). Inspecting Figure 2d we see the instrumental variable assumption does sharpen the limiting bounds of $\Psi$ slightly, however if we consider inference using 90\% or 95\% equal-tailed credible intervals we can see intervals under the instrumental variable assumption will actually be wider than with no assumptions. This is because there is perfect correlation between $\omega$'s in the $Z=0$ and $Z=1$ cells when we assume $Z$ is an instrument, resulting in more frequent extreme probability values relative to the no assumption case, and increasing posterior variance.

Further assumptions on the missing data are needed in these situations to sharpen posterior inference. Many assumptions are possible, and reasonable ones will depend on the specific problem of interest. Here we will discuss another general assumption on the missing data mechanism that we feel will be broadly applicable. When nonignorable missingness is present we generally do not know the magnitude of the bias induced by the missing data mechanism (i.e. the magnitude of $P(Y=1|R=0,X=x,Z=z) - P(Y=1|R=1,X=x,Z=z)$), but we may know the direction of this bias. For example, we may be certain that the outcome is more prevalent among those who are missing, $P(Y=1|R=0,X=x,Z=z) > P(Y=1|R=1,X=x,Z=z) \hspace{0.1cm} \forall \hspace{0.05cm} x,z$, which we describe as a 'positive bias', or we may have some prior belief that this is the case. 

We can incorporate these prior beliefs similarly to how we included instrumental variable assumptions in Section 2.1. For example, if we assume $Z$ is an instrumental variable and there is a positive bias in the missing data our joint prior becomes

\begin{align}
f(q_{0,0},q_{0,1},q_{1,0},q_{1,1},\omega_{0,0},\omega_{0,1},\omega_{1,0},\omega_{1,1}) \propto & f(q_{0,0} ; \bm{\alpha}_{0,0}) f(q_{0,1} ; \bm{\alpha}_{0,1})f(q_{1,0} ; \bm{\alpha}_{1,0})f(q_{1,1} ; \bm{\alpha}_{1,1}) \nonumber \\
&I_{\left ( \frac{q_{0,0,1,1}}{(1-q_{0,0,0,\boldsymbol{\cdot}})},1 \right)}(\omega_{0,0})I_{\left (\frac{q_{0,1,1,1}}{(1-q_{0,1,0,\boldsymbol{\cdot}})},1 \right)}(\omega_{0,1}) \\
&I_{\left (\frac{q_{1,0,1,1}}{(1-q_{1,0,0,\boldsymbol{\cdot}})},1 \right)}(\omega_{1,0})I_{\left (\frac{q_{1,1,1,1}}{(1-q_{1,1,0,\boldsymbol{\cdot}})},1 \right)}(\omega_{1,1}) \nonumber. 
\end{align}

We can reparametrize as in (4) to get a joint prior in the $X=0$ stratum of 

\begin{align}
f(q_{0,0},q_{0,1},\omega_{0,1}^*, \omega_{0,0}^* = 0) &= f(q_{0,0},q_{0,1},\omega_{0,1}^*| \omega_{0,0}^* = 0)f(\omega_{0,0}^* = 0) \propto \nonumber \\
f_{q_{0,0}}f_{q_{0,1}}\frac{1}{q_{0,0,0,\boldsymbol{\cdot}}}\frac{1}{q_{0,1,0,\boldsymbol{\cdot}}}&I_{\left(max \left(\frac{\frac{q_{0,0,1,1}q_{0,0,0,\boldsymbol{\cdot}}}{(1-q_{0,0,0,\boldsymbol{\cdot}})} + q_{0,0,1,1} - q_{0,1,1,1}}{q_{0,1,0,\boldsymbol{\cdot}}},\frac{q_{0,1,1,1}}{(1 - q_{0,1,0,\boldsymbol{\cdot}})} \right),min \left(1,\frac{q_{0,0,1,1} - q_{0,1,1,1} + q_{0,0,0,\boldsymbol{\cdot}}}{q_{0,1,0,\boldsymbol{\cdot}}}\right) \right)}(\omega_{0,1})\nonumber \\ 
&\delta_{\omega_{0,0}}\left( \frac{q_{0,1,1,1} - q_{0,0,1,1} + q_{0,1,0,\boldsymbol{\cdot}}\omega_{0,1}}{q_{0,0,0,\boldsymbol{\cdot}}} \right),
\end{align}

and we will refer to (7) as the "Sat-OR-1-pos" model. The conditional priors on $\omega_{x,z}$ given $q_{x,z}$ become

\begin{align*}
[\omega_{0,1}|{q_{x,z}} ] &\sim Unif \Bigg (max \left (\frac{\frac{q_{0,0,1,1}q_{0,0,0,\boldsymbol{\cdot}}}{(1-q_{0,0,0,\boldsymbol{\cdot}})} + q_{0,0,1,1} - q_{0,1,1,1}}{q_{0,1,0,\boldsymbol{\cdot}}}, \frac{q_{0,0,1,1} - q_{0,1,1,1}}{q_{0,1,0,\boldsymbol{\cdot}}} \right), \\
& \hspace{1.75cm} min \left (1, \frac{q_{0,0,1,1} - q_{0,1,1,1} + q_{0,0,0,\boldsymbol{\cdot}}}{q_{0,1,0,\boldsymbol{\cdot}}}\right ) \Bigg ) \\
[\omega_{1,1}|{q_{x,z}} ] &\sim Unif \Bigg (max \left (\frac{\frac{q_{1,0,1,1}q_{1,0,0,\boldsymbol{\cdot}}}{(1-q_{1,0,0,\boldsymbol{\cdot}})} + q_{1,0,1,1} - q_{1,1,1,1}}{q_{1,1,0,\boldsymbol{\cdot}}}, \frac{q_{1,0,1,1} - q_{1,1,1,1}}{q_{1,1,0,\boldsymbol{\cdot}}} \right), \\
& \hspace{1.75cm} min \left (1, \frac{q_{0,0,1,1} - q_{0,1,1,1} + q_{0,0,0,\boldsymbol{\cdot}}}{q_{0,1,0,\boldsymbol{\cdot}}}\right ) \Bigg ) \\
[\omega_{0,0}|{q_{x,z}} ]&\sim \delta_{\omega_{0,0}}\left( \frac{q_{0,1,1,1} - q_{0,0,1,1} + q_{0,1,0,\boldsymbol{\cdot}}\omega_{0,1}}{q_{0,0,0,\boldsymbol{\cdot}}} \right) \\
[\omega_{1,0}|{q_{x,z}} ] &\sim \delta_{\omega_{1,0}}\left( \frac{q_{1,1,1,1} - q_{1,0,1,1} + q_{1,1,0,\boldsymbol{\cdot}}\omega_{1,1}}{q_{1,0,0,\boldsymbol{\cdot}}} \right).
\end{align*}

Comparing these intervals to those given with just an instrumental assumption, Figure 2c and 2d show that the positive missing data bias assumption increases the lower bound of $P(Y=1|X=1)$ and as a result greatly narrows the nonparametric bounds and credible intervals of $\Psi$ in example 2. 

We can again characterize our prior assumption about a positively biased missing data mechanism using a probability distribution as in the Sat-OR-normal model. Our assumption of positive bias assumes $\omega_{x,z} > \frac{q_{x,z,1,1}}{(1-q_{x,z,0,\boldsymbol{\cdot}})}$. Transforming this difference to a range of $(0,1)$ we can characterize our prior belief, for example, as $h(\omega_{x,z}| q_{x,z}) = \frac{\left( \omega_{x,z} - \frac{q_{x,z,1,1}}{(1-q_{x,z,0,\boldsymbol{\cdot}})} + 1 \right)}{2} \sim beta(a,b)$, with $a > b$. These beta distributions can be combined with a prior assumptions on $OR(Y,Z|X)$, see the Appendix for details on the joint prior. We will refer to this as the "Sat-OR-normal-beta" model. 

\subsection{Adjusting for confounding}

Our problem setting described in Section 1.2 is primarily applicable to RCTs where no adjustment for confounding is necessary. However, it can be extended to allow for inclusion of discrete confounding variables if needed. To accomplish this for a set of discrete confounders $C$ of length $p$, we can define separate priors over all unique combinations of $C=c$. We can then estimate our parameter of interest $\Psi$ as the weighted average

\begin{equation*}
\Psi = \sum_c \begin{aligned}
&\left(
\begin{aligned}
& \left( (q_{1,1,1,1,c} + q_{1,1,0,\boldsymbol{\cdot},c}\omega_{1,1,c})q_z +  
         (q_{1,0,1,1,c} + q_{1,0,0,\boldsymbol{\cdot}}\omega_{1,0,c})(1-q_z) \right) - \\
         & \left( (q_{0,1,1,1,c} + q_{0,1,0,\boldsymbol{\cdot},c}\omega_{0,1,c})q_z +  
         (q_{0,0,1,1,c} + q_{0,0,0,\boldsymbol{\cdot},c}\omega_{0,0,c})(1-q_z) \right)
\end{aligned}
\right)\theta_c
\end{aligned}
\end{equation*}

where $\theta_c = P(C = c)$. Assuming $\theta_c$ is independent of $q_{x,z,c}$ and $\omega_{x,z,c}$, priors for $q_{x,z,c}$ and $\omega_{x,z,c}$ are set as in Section 2.1 or 2.2. Taking a $Dir(\kappa, ..., \kappa)$ prior on $\theta_C$ gives us our full model. Unfortunately this setup will quickly become problematic as the number of confounders increases, as it requires separate modelling over all $2^p$ unique confounder combinations. There are ways to try to sidestep this problem (e.g. Daly-Grafstein and Gustafson 2022), though the method is probably best suited for situations with only a small number of discrete confounders. 

\section{Model computation and inference}

In this section we discuss computation of the posterior distributions implied by the priors above. In the case of our Sat model with prior given in (1), computation can proceed using standard conjugate updating. Let $n_{x,z} = (n_{x,z,1,0}, n_{x,z,1,1}, n_{x,z,0,\boldsymbol{\cdot}})$ denote the observed counts for combinations of $Y$ and $R$ when $X=x$ and $Z=z$. Our Dirichlet priors in (1) for $q_{x,z}$ can be updated in conjugate fashion resulting in a posterior

\begin{align}
f(q_{0,0},q_{0,1},q_{1,0},q_{1,1},\omega_{0,0},\omega_{0,1},\omega_{1,0},\omega_{1,1}) = & f_{q_{0,0}}(\bm{\alpha}_{0,0} + n_{0,0}) f_{q_{0,1}}(\bm{\alpha}_{0,1} + n_{0,1}) \nonumber \\
&f_{q_{1,0}}(\bm{\alpha}_{1,0} + n_{1,0}) f_{q_{1,1}}(\bm{\alpha}_{1,1} + n_{1,1})\\
&I_{(0,1)}(\omega_{0,0})I_{(0,1)}(\omega_{0,1})I_{(0,1)}(\omega_{1,0})I_{(0,1)}(\omega_{1,1}), \nonumber
\end{align}

which can be computed using independent Monte Carlo draws from each component Dirichlet and Uniform distribution.

When we introduce the assumption of an instrumental variable our $q_{x,z}$ and $\omega_{x,z}$ parameters are no longer independent. For the Sat-OR-1 model the prior distribution on $\omega_{x,z}$ in (2) depends on the values of $q_{x,z}$, as seen by the reparameterization in (4). To account for this we propose a rejection sampling algorithm based on the posterior in (8). Without loss of generality consider the parameters in (2) corresponding to the $X=0$ stratum, as even with our instrumental variable assumption parameters in the $X=1$ and $X=0$ strata are independent. We can write our joint prior (2) as $f(q_{0,1}, q_{0,0}, \omega_{0,0}, \omega_{0,1}) = f(\omega_{0,0}|\omega_{0,1},q_{0,1}, q_{0,0})f(\omega_{0,1}|q_{0,1}, q_{0,0})f(q_{0,1}, q_{0,0})$ where these distributions correspond to the Dirac, Uniform, and Dirichlet distributions in (4), respectively. In the posterior the Dirichlets can again be updated using conjugacy and we can proceed with a rejection sampling algorithm by using a proposal distribution that ignores the instrumental variable assumption that defines $f(\omega_{0,1}|q_{0,1}, q_{0,0})$. Specifically, we sample independently from two $Dir(\alpha_1 + n_{x,z,1,0}, \alpha_2 + n_{x,z,1,1}, \alpha_3 + n_{x,z,0,\boldsymbol{\cdot}}- 1)$ distributions, sample $\omega_{0,1}$ from a $Unif(0,1)$, and set $\omega_{0,0} =  \frac{q_{0,1,1,1} - q_{0,0,1,1} + q_{0,1,0,\boldsymbol{\cdot}}\omega_{0,1}}{q_{0,0,0,\boldsymbol{\cdot}}}$. We accept samples with probability 1 if 

$$\omega_{0,1} \in \left[max \left(0,\frac{q_{0,0,1,1} - q_{0,1,1,1}}{q_{0,1,0,\boldsymbol{\cdot}}} \right), min \left(1,\frac{q_{0,0,1,1} - q_{0,1,1,1} + q_{0,0,0,\boldsymbol{\cdot}}}{q_{0,1,0,\boldsymbol{\cdot}}}\right) \right]$$ 

as defined by the conditional uniform prior in (4), otherwise we reject. We sample similarly for parameters in the $X=1$ stratum and combine estimates as in (5) to compute the posterior of $\Psi$. Sampling from the Sat-OR-1-pos model proceeds similarly.

We can also use rejection sampling when sampling from the posterior of the Sat-OR-thresh model where we assume each OR is in some interval around 1. Because this constraint does not have an area of $0$ in the parameter space when $t_h > t_l$, we do not need to reparametrize the distribution as in (4). We simply sample independently from the marginal distributions in (8) and accept samples with probability 1 if the OR falls within the bounds $(t_l, t_h)$, otherwise we reject. Sampling from the Sat-OR-normal and Sat-OR-normal-beta models proceeds similarly.

Constructing our rejection sampler using the Sat model as the proposal gives us a way to calculate a Bayes factor that quantifies the evidence for these assumptions as part of the sampling process. If we define all our parameters $\theta = \{q_{x,z}, \omega_{x,z}\}$, we can view any of the previous models with missing data assumptions as extensions of the Sat model with a restriction on $\theta$ to some set $\mathcal{A}$. Let $D = \{X, Z, Y_{(R = 1)}\}$ define our observed data. If our posterior for the Sat model given in (8) is $f(\theta|D)$, then the posterior for our model with restrictions can be taken to be proportional to $f(\theta|D)I_\mathcal{A}(\theta)$. If we want to compare the evidence for the Sat model compared to a model with missing data assumptions, we can write the Bayes factor as

\begin{align}
BF &= \frac{\int f(\theta)L(D|\theta)d\theta}{\int f(\theta)I_\mathcal{A}(\theta)L(D|\theta)d\theta/\int f(\tilde{\theta})I_\mathcal{A}(\tilde{\theta})d\tilde{\theta})} \\
&= \frac{1}{AR} \int f(\theta)I_\mathcal{A}(\theta)d\theta, \nonumber
\end{align}

where $AR$ is the acceptance rate of our rejection sampler given by $\int_\mathcal{A} L(D|\theta)f(\theta)/\int f(\tilde{\theta})L(D|\tilde{\theta})d\tilde{\theta}$. Additionally, $\int f(\theta)I_\mathcal{A}(\theta)d\theta$ is simply the acceptance rate of our same rejection sampler taken with respect to the priors over the Sat and missing data models, rather than the posteriors. Thus the Bayes factor comparing models with and without missing data assumptions can be computed using the ratio of acceptance rates of our rejection algorithm for the priors over the posteriors. We will use this result in the simulations in Section 4 to quantify the evidence for missing data assumptions present in the data.  

Finally, we consider inference in the limit with infinite data. In this case $\Psi$ is only partially identified, characterized by a limiting posterior distribution (LPD) rather than a point mass. In the limit parameters dependent on the observed data approach their true values $q_{x,z}^\dagger$, and the LPD is determined by the conditional priors $f(\omega_{x,z}|q_{x,z} = q_{x,z}^\dagger)$. As we approach the limit it is possible for the acceptance rate of our rejection algorithm to go to 0. Consider the Sat-OR-1 model with an instrumental variable assumption given in (4). The conditional prior $f(\omega_{0,1}|q_{x,z} = q_{x,z}^\dagger)$ is not defined when $q_{0,1,1,1}^\dagger - q_{0,0,1,1}^\dagger > q_{0,0,\boldsymbol{\cdot}}^\dagger$ or when $q_{0,0,1,1}^\dagger - q_{0,1,1,1}^\dagger > q_{0,1,0,\boldsymbol{\cdot}}^\dagger$. In these cases it is not possible for $Z$ to be an instrument because the observed probabilities $q_{x,z}^\dagger$ are too far apart to equalize $P(Y=1|X=0, Z=1)$ and $P(Y=1|X=0,Z=0)$ using the missing data. Thus while instrumental and other missing data assumptions are not empirically verifiable given the data, in some cases it is possible to falsify these assumptions. See Balke and Pearl (1997), Mattei et al. (2014), and Swanson et al. (2019) for further discussion of instrumental variable bounds. 

\section{Simulation studies}

\subsection{Simulation design}

We aim to evaluate our proposed method using the scenario described in Section 1.1 with two simulation studies. To do this we will compute the frequentist coverage of the credible interval for $\Psi$ under different missing data assumptions, and compare them to a Heckman selection model. As a reminder, in this scenario we have a single binary treatment variable $X$, a potential binary instrument $Z$, binary outcome $Y$ which is partially missing, and we are interested in the effect of $X$ on $Y$ defined as $\Psi = P(Y|X=1) - P(Y|X=0)$. 

In some settings, when we include certain missing data assumptions the efficiency of our rejection sampling algorithm may decrease. Specifically, if the situation described in the previous section occurs where in the LPD our conditional prior $f(\omega_{0,1}|q_{x,z} = q_{x,z}^\dagger)$ is not defined, the acceptance rate of our algorithm will go to 0. While this will not happen in finite sample settings, we can have situations where the acceptance rate becomes prohibitively low. This is actually a feature of our method, allowing us to quantify the evidence of our missing data assumptions given by the data, but makes frequentist coverage comparisons difficult in some simulation settings. Therefore we will split our simulation study into two parts. In the first data are generated with the instrumental variable assumption and positive missing data bias assumption holding. In this case the acceptance rate of all models described previously will be relatively high so posteriors can be computed in all case and frequentist comparisons between methods can be made. In the second part one or both of the instrumental variable assumption or positive missing data bias assumption will be false. This can result in very inefficient rejection samplers in some cases. In cases where the acceptance rate is less than some minimum threshold we will consider the posterior as having failed to compute. We will make frequentist comparisons of our credible intervals for the subset of simulation runs where posteriors are successfully computed, as well as compare the proportion of runs that computed and evidence for the missing data assumptions. In each case we will compare the Sat, Sat-OR-1, Sat-OR-thresh, Sat-OR-normal, Sat-OR-1-pos, Sat-OR-normal-beta, Heckman selection model, and an oracle model where no data are missing. Posteriors can be computed via conjugate updating for the saturated model with no assumptions and the oracle. We use rejection sampling for all models with missing data assumptions as described in Section 3. For the Heckman selection model we approximate the posterior distribution using a Gibbs sampler (van Hasselt 2011). See Appendix A.3 for the full Gibbs sampling algorithm. 

\subsection{Simulation 1: missing data assumptions hold}

Our first simulation study is conducted with a true instrumental variable and a positive missing data bias. We generate data using two methods, in the first data are simulated using the Heckman selection model. We set $\rho = -0.5$ so the positive bias holds, $P(X=1) = P(Z=1) = 0.5$, $\beta_0 = -0.5$, and $\beta_1 = 0.75$. We use two levels of missingness for $Y$, 0.2 and 0.4, and $\bm{\gamma}$ values are set such that these levels of missingness are achieved. In the second method $q_{x,z}$ and $\omega_{x,z}$ are generated using the Sat-OR-1-pos prior such that the parametric assumptions of the Heckman selection model are not true, but the positive bias and instrumental assumptions hold. Here we generate $p_{x,1}$ with a given level of missingness, and $p_{x,0}$ are set such that (3) holds. We repeat generations until $q_{x,z,\boldsymbol{\cdot}}\omega_{x,z}/q_{x,z,0,\boldsymbol{\cdot}} > q_{x,z,1,1}/(q_{x,z,1,1} + q_{x,z,1,0})$ holds for all $X,Z$ such that the positive data bias assumption is satisfied. In all cases we set $n=1000$ for the simulated dataset. Parameter values are set as follows: $\bm{\alpha} = (1,1,2)$, $t_l = 2/3$, $t_h = 3/2$, $\sigma = 0.4$, and $(a, b) = (4,2)$. In total we have 2 data generating processes (DGPs) and 2 levels of missingness for a total of 4 simulations settings. We repeat each 200 times, computing the posterior for $\Psi$ for each model and recording the width of the 90\% equal-tailed credible interval, whether it contained the true value of $\Psi$, and the acceptance ratio of the rejection sampler where applicable. 

\begin{table}[!t]
    \centering
    \caption{Mean acceptance rate, (geometric) mean Bayes factor, coverage, and mean 90\% credible interval width over 200 runs of simulation 1. The "Heckman" model assumes a Bayesian Heckman selection model, and the "Oracle" model fits a saturated model without missing data. Data are simulated either using a Heckman DGP or a saturated DGP with instrumental and positive bias assumptions. All Bayes factor compare the posterior odds of the Sat model to the model of interest.}
    \vspace{0.3cm}
    \label{tab:mytable}
    \renewcommand{\arraystretch}{1.12} 
    \begin{tabular}{|c|c|c|c|c|c|}
        \hline
        \multirow{2}{*}{\textbf{Simulation}} & \multirow{2}{*}{\textbf{Model}} & \multicolumn{4}{c|}{\textbf{Results}} \\
        \cline{3-6}
        & & \textbf{AR} & \textbf{BF} & \textbf{Coverage} & \textbf{Interval width}  \\
        \hline
        \multirow{8}{*} & Sat & & & 1 & 0.24 \\
        & Sat-OR-1 & 0.81 & 0.21 & 0.99 & 0.20 \\
        & Sat-OR-thresh & 0.29 & 0.12 & 1 & 0.22 \\
        Heckman DGP & Sat-OR-normal & 0.35 & 0.15 & 1 & 0.22 \\
        $P(R=0) = 0.2$ & Sat-OR-1-pos & 0.20 & 0.13 & 0.98 & 0.17\\
        & Sat-OR-normal-beta & 0.07 & 0.07 & 0.99 & 0.20\\
        & Heckman & & & 0.88 & 0.12 \\
        & Oracle & & & 0.88 & 0.10 \\
        \hline
        \multirow{8}{*} & Sat & & & 1 & 0.39  \\
        &  Sat-OR-1 & 0.96 & 0.17 & 1 & 0.46 \\
        & Sat-OR-thresh & 0.14 & 0.26 & 1 & 0.44 \\
        Heckman DGP & Sat-OR-normal & 0.19 & 0.28 & 1 & 0.43 \\
        $P(R=0) = 0.4$  &  Sat-OR-1-pos & 0.37 & 0.06 & 1 & 0.33 \\
        & Sat-OR-normal-beta & 0.05 & 0.10 & 1 & 0.36 \\
        & Heckman & & & 0.88 & 0.14 \\
        & Oracle & & & 0.89 & 0.10\\
        \hline
        \multirow{8}{*} & Sat & & & 0.98 & 0.21 \\
        & Sat-OR-1 & 0.50 & 0.34 & 0.99 & 0.23 \\
        & Sat-OR-thresh & 0.37 & 0.10 & 0.99 & 0.22\\
        Saturated DGP & Sat-OR-normal & 0.42 & 0.13 & 0.99 & 0.22 \\
        $P(R=0) = 0.2$ &  Sat-OR-1-pos & 0.14 & 0.24 & 0.93 & 0.18\\
        & Sat-OR-normal-beta & 0.06 & 0.08 & 0.98 & 0.19\\
        & Heckman & & & 0.75 & 0.11 \\
        & Oracle & & & 0.91 & 0.10\\
        \hline
        \multirow{8}{*} & Sat  & & & 1 & 0.39  \\
        &  Sat-OR-1 & 0.63 & 0.26 & 1 & 0.45 \\
        & Sat-OR-thresh & 0.13 & 0.28 & 1 & 0.43\\
        Saturated DGP & Sat-OR-normal & 0.18 & 0.30 & 1 & 0.43\\
        $P(R=0) = 0.4$  &  Sat-OR-1-pos & 0.24 & 0.13 & 0.94 & 0.34\\
        & Sat-OR-normal-beta & 0.03 & 0.19 & 0.99 & 0.35\\
        & Heckman & & & 0.60 & 0.12 \\
        & Oracle & & & 0.90 & 0.10  \\
        \hline
    \end{tabular}
\end{table}

Table 2 shows the results our of first simulation study. Coverage are above nominal levels for all saturated models in each simulation setting, but are below nominal levels for the Heckman model when the parametric assumptions of the model are not satisfied. In terms of interval width, we can see many of the saturated models with just the instrumental variable assumption have on average wider 90\% credible intervals compared to the model with no assumptions, similar to what we saw in Section 2.2. Only when adding in the assumption about the direction of missing data bias do we see a reduction of average interval width compared to the model with no assumptions.

We can use the acceptance rates of the rejection samplers computing the prior over the posterior to compute the posterior odds of our missing data assumptions via the Bayes factor. Without any data, the prior acceptance rates for the Sat-OR-1, Sat-OR-thresh, Sat-OR-normal, Sat-OR-1-pos, and Sat-OR-normal-beta are 0.160, 0.035, 0.051, 0.023, and 0.005, respectively. Note the Sat-OR-1 model uses a slightly different rejection sampling algorithm than the Sat-OR-thresh or Sat-OR-normal models, which is why prior acceptance rates differ so greatly between the models. However, maintaining a consistent rejection sampling procedure between the prior and posterior allows us to estimate the Bayes factor in all cases. Focusing on the setting where data are generated using a Heckman DGP with a missingness of 0.2, the Bayes factor results indicate that there is evidence in most of the datasets generated under these parameter values for both the instrumental variable assumption and positive data bias assumption, as mean Bayes factors are below 1. Additionally, as we add more assumptions that are supported by the data, Bayes factors decrease.

\subsection{Simulation 2: missing data assumptions do not hold}

Here we repeat the simulation as above except we generate data with either the instrumental variable assumption, positive data bias assumption, or both not holding. When using a Heckman DGP we violate the instrumental variable assumption by including an extra parameter $\beta_2 = 1.5$ into the model such that the outcome equation for $Y$ becomes $P(Y_i=1|X_i, Z_i) = \Phi(\beta_0 + \beta_1 X_i + \beta_2 Z_i - \beta_2X_iZ_i)$. To violate the positive data bias assumption we set $\rho = 0.5$. With the saturated DGP, we generate $p_{x,z}$ using the prior in (1) to guarantee the $OR=1$ assumption does not hold, and repeat generations until $q_{x,z,\boldsymbol{\cdot}}\omega_{x,z}/q_{x,z,0,\boldsymbol{\cdot}} < q_{x,z,1,1}/(q_{x,z,1,1} + q_{x,z,1,0})$ for all $X,Z$ to ensure the positive data bias assumption does not hold. We again use two levels of missingness for $Y$, 0.2 and 0.4, resulting in 12 total simulation settings, each repeated 200 times. We consider posteriors as having failed to compute when acceptance rates result in Bayes factors larger than 10 in favour of the saturated model with no assumptions. For example, the prior acceptance rate of the Sat-OR-1 model is 0.16. We require 1000 samples from our posterior, so if after $1000/0.016 = 62500$ samples from our rejection sampler we have not accepted at least 1000 we consider that posterior computation to have failed.

\begin{table}[!t]
    \centering
    \caption{Mean acceptance rate, (geometric) mean Bayes factor, proportion of posteriors computed, coverage, and mean 90\% credible interval width over 200 runs from simulation 2. Note coverage and credible intervals are calculated on the subset of runs where the posterior is computed successfully. The "Heckman" model assumes a Bayesian Heckman selection model, and the "Oracle" model fits a saturated model without missing data. Data are simulated either using a Heckman DGP or a saturated DGP with one or both of the instrumental variable and positive data bias assumptions violated. All Bayes factors compare the posterior odds of the Sat model to the model of interest.}
    \vspace{0.3cm}
    \label{tab:mytable}
    \renewcommand{\arraystretch}{1.12} 
    \begin{tabular}{|c|c|c|c|c||c|c|}
        \hline
        \multirow{2}{*}{\textbf{Simulation}} & \multirow{2}{*}{\textbf{Model}} & \multicolumn{5}{c|}{\textbf{Results}} \\
        \cline{3-7}
        & & \textbf{AR} & \textbf{BF} & \textbf{Computed} & \textbf{Coverage} & \textbf{Interval width} \\
        \hline
        \multirow{5}{*} & Sat & &  &  & 0.98 & 0.21   \\
        Saturated DGP & Sat-OR-1 & 1.8e-03 & 798 & 0.23 & 0.89 & 0.14  \\
        $P(R=0) = 0.2$ & Sat-OR-1-pos & 2.5e-7 & 5750 & 0 & NA & NA  \\
        $OR \neq 1$, pos bias & Heckman & &  &  & 0.73 & 0.10   \\
        & Oracle & &  &  & 0.88 & 0.10  \\
        \hline
        \multirow{5}{*} & Sat &   &  &  & 1 & 0.39\\
        Saturated DGP & Sat-OR-1 & 0.24 & 0.72 & 1 & 0.91 & 0.31 \\
        $P(R=0) = 0.4$ & Sat-OR-1-pos  & 0.008 & 93.7 & 0.26 & 0.82 & 0.36 \\
        $OR \neq 1$, pos bias & Heckman  & &  &  & 0.41 & 0.11 \\
        & Oracle &  &  &  & 0.87 & 0.10 \\
        \hline
        \multirow{5}{*} & Sat &  &  &  & 0.98 & 0.21 \\
        Saturated DGP & Sat-OR-1 & 4.5e-03 & 1060 & 0.22 & 0.67 & 0.14 \\
        $P(R=0) = 0.2$ & Sat-OR-1-pos & 2.2e-06 & 261 & 0 & NA & NA \\
        $OR \neq 1$, neg bias & Heckman & &  &  & 0.68 & 0.11  \\
        & Oracle & &  &  & 0.89 & 0.10  \\
        \hline
        \multirow{5}{*} & Sat &  &  &  & 1 & 0.39 \\
        Saturated DGP & Sat-OR-1 & 0.27 & 0.69 & 1 & 0.93 & 0.32 \\
        $P(R=0) = 0.4$ & Sat-OR-1-pos & 0.004 & 92.6 & 0.28 & 0.91 & 0.28 \\
        $OR \neq 1$, neg bias & Heckman &  &  &  & 0.38 & 0.11 \\
        & Oracle &  &  &  & 0.90 & 0.10 \\
        \hline
        \multirow{5}{*} & Sat &  & &  & 0.95 & 0.21  \\
        Saturated DGP & Sat-OR-1 & 0.20 & 1.43 & 0.97 & 0.95 & 0.19 \\
        $P(R=0) = 0.2$ & Sat-OR-1-pos & 0.026 & 19.6 & 0.43 & 0.86 & 0.18 \\
        $OR = 1$, neg bias & Heckman & &  &  & 0.58 & 0.11  \\
        & Oracle &  &  &  & 0.92 & 0.10 \\
        \hline
        \multirow{5}{*} & Sat & &  &  & 0.87 & 0.38  \\
        Saturated DGP & Sat-OR-1 & 0.30 & 0.85 & 1 & 0.88 & 0.34 \\
        $P(R=0) = 0.4$ & Sat-OR-1-pos & 0.075 & 2.51 & 0.46 & 0.59 & 0.35 \\
        $OR = 1$, neg bias & Heckman &  &  &  & 0.25 & 0.12 \\
        & Oracle & &  &  & 0.90 & 0.10  \\
        \hline
    \end{tabular}
\end{table}

In Table 3 we present a subset of the results from simulation 2, see Appendix A.4 for the full results. Coverage and interval width results are taken only for the posteriors that are successfully computed for a given model. The coverage results given for models where few posteriors are computed are likely overly optimistic. When posteriors are not computed the data give stronger evidence against our assumptions, meaning the data suggest odds ratios are likely further from 1, or the missing data bias is more likely negative, rather than positive. In these cases it is likely coverage would have been lower had posteriors been computed. 

We can see in cases where prior assumptions are violated coverage results are lower. When the odds ratio assumption is violated and missingness is low acceptance rates for the Sat-OR-1-pos model are quite low, and we fail to compute the posterior in all cases. The failure of our rejection sampler does still give us information, however. We can see this by looking at the average acceptance rate of the rejection samplers for each of the simulation settings and the corresponding Bayes factors. For example, looking at the Bayes factor for the Sat-OR-1 model, we can see a larger proportion of posteriors are not computed and acceptance rates are lower when the instrumental variable assumption is violated compared to when it is not. Comparing simulations with different missingness rates, we can see Bayes factors for models with violated assumptions are higher when missingness is low. In these cases \textit{a posteriori} we are more confident that model assumptions are violated, compared to similar situations with a higher proportion missingness. Finally, in cases where Bayes factors are high the corresponding coverages tend to be lower. This is an advantage of our method over the Heckman model where we get no information about the plausibility of model assumptions. 

\clearpage

\section{Example: New leaf pilot study}

Here we give an example of our method applied to the New Leaf Pilot study (NLP) conducted by the Foundation for Social Change, a RCT examining the effects of a one-time cash transfer on individuals experiencing homelessness in Vancouver, BC (Dwyer et al. 2021). This study provided a \$7,500 cash transfer to 50 individuals experiencing homelessness, with another 65 serving as controls. Outcomes of interest, including cognitive function, life satisfaction, spending, employment, and housing status were taken at baseline, then again at 1, 3, 6, 9, and 12 month follow up surveys conducted from 2016 to 2020. We analyze a subset of the data to answer a simpler question: what was the effect of the one-time cash transfer on participant housing status 3 months after baseline? We take $X_i$ to be the condition of participant $i$, either cash or control, and $Y_{i3}$ to be the housing status of that participant at the 3 month follow up survey, with $Y_{i3} = 1$ indicating the participant is in stable housing, and $Y_{i3} = 0$ indicating the participant is homeless. Because all participants are homeless at baseline, our parameter of interest is $\Psi = P(Y_{3}|X=1) - P(Y_{3}|X=0)$.   

All conditions $X$ are known, however many outcomes $Y_{i3}$ are missing in the data due to participant dropout and missed surveys. A total of 30 out of the 115 participants have missing housing statuses at the 3 month follow up. We also expect this missingness to be nonignorable - we expect participants who have found stable housing and no longer need the supports provided by the study to be more likely to be missing. While there is not a clear instrumental variable to use in our analysis, for this example we will take the participant age, dichotomized as greater than or less than 40 years old, to be our instrument $Z$. Because we do not expect $Z$ to be a perfect instrument, and have some uncertainty about the direction of missing data bias, we use probabilistic priors on $OR(Y,Z|X)$ and $h(\omega_{x,z}|q_{x,z})$ as in the Sat-OR-normal-beta model. We set $\bm{\alpha} = (1,1,2)$, $\sigma = 0.4$, $a = 5$, and $b=2$. We compare this model to our Sat model that makes no assumptions about the variable $Z$ and missing data bias, a Heckman selection model as described in Section 1.2, and a model where we assume $Y$ are MAR given $X$ and $Z$. Our parameter $\Psi$ is computed as in (5), and we give $q_z$ a $beta(1,1)$ prior. 

Posterior credible intervals for each of the four approaches are given in Figure 3. In all cases 95\% credible intervals include 0, however 80\% credible intervals for $\Psi$ under the Heckman selection, MAR, and Sat-OR-normal-beta models do not. We find the posterior probability of the one-time cash transfer increasing the probability of finding stable housing after 3 months to be 0.94, 0.82, 0.96, and 0.92, using the MAR, Sat, Heckman, and Sat-OR-normal-beta models, respectively. Incorporating prior assumptions on $Z$ and the direction of missing data bias into our saturated model partially sharpened inference, without needing to make the assumptions necessary in the Heckman or MAR models. 

\begin{figure}[!t]
    \centering
    {\includegraphics[width=0.8\textwidth]{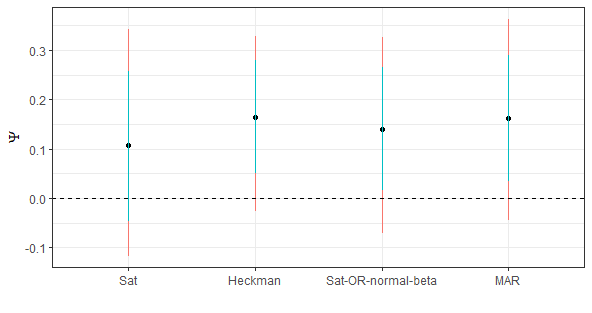}}
    \caption{Posterior credible intervals for $\Psi$ under our four models. Dots represent the posterior mean, blue and red lines the 80\% and 95\% equal-tailed credible intervals, respectively. The Heckman model is computed as described in Section 1.2, and the MAR model is computed using a saturated model assuming $Y \perp R|X,Z$, allowing us to ignore cases where $Y$ is missing.}
    \label{fig:combined}
\end{figure}

\section{Discussion}

In this paper we proposed a new method for handling discrete nonignorable missing outcome data. We juxtapose our method with a Heckman selection model, where we trade-off strong parametric assumptions that allow for identifiability but possible model misspecification for the relaxation of $\textit{a priori}$ parametric assumptions and partial identification. While we focused on missing outcomes, nonignorable missingness in the covariates can likely be handled similarly. For a set of covariates $C$ and missingness in both the outcomes and covariates, the dimension of our missingness indicator $R$ would increase, and we can specify a joint model and missing data assumptions for the combined outcome, covariate, and missingness indicators.  

Throughout the paper we used a Dirichlet prior $[q_{x,z}] \sim Dir(\bm{\alpha})$, taking $\bm{\alpha} = (\alpha_1, \alpha_2, \alpha_3 + \alpha_4) = (1,1,2)$ for all models, and assumed $\alpha_3 = \alpha_4$. With reasonably sized datasets where cell counts in each combination of $X$ and $Z$ (and potentially some covariates $C$) are large these priors are unlikely to have a large effect on the inference of $\Psi$. However, we also use the prior acceptance rate as part of our calculation of the Bayes Factor in (9), and changes in $[q_{x,z}]$ can result in large changes in this rate. Here $q_{x,z}$ parameterizes the joint multinomial distribution of $Y$ and $R$, with categories $(Y=0, R=1)$, $(Y=1, R=1)$, and $(R=0)$, respectively. Taking our hyperparameter $\bm{\alpha} = (1,1,2)$ and increasing the $\textit{a priori}$ probability of $R=0$, say for example by taking $\bm{\alpha} = (1,1,100)$ (i.e. assuming $\alpha_3 = \alpha_4 = 50$), greatly increases the prior acceptance rate for all models.

\begin{figure}[!t]
    \centering
    \subfloat[]{\includegraphics[width=0.5\textwidth]{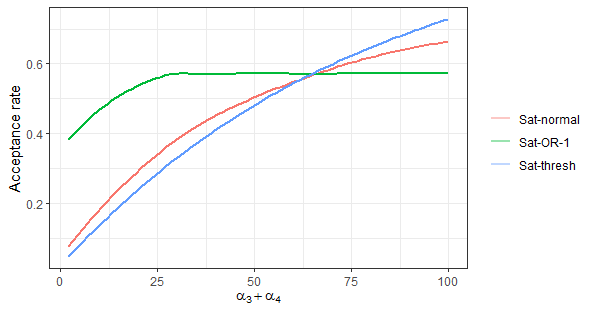}}\hfill
    \subfloat[]{\includegraphics[width=0.5\textwidth]{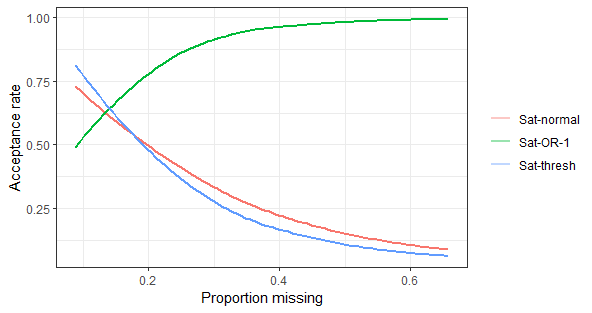}}
    \caption{Figure (a) shows the prior acceptance rate of our method using different priors over varying values of $\alpha_3 + \alpha_4$ with $\alpha_1 = \alpha_2 = 1$ and $\alpha_3 = \alpha_4$. Figure (b) shows the posterior acceptance rate using the same hyperparameter $\bm{\alpha} = (1,1,2)$ in all models and observing a dataset with $n=1000$, where an increasingly large proportion of the data is missing.}
    \label{fig:combined}
\end{figure}

We can see this phenomenon in Figure 4. The left panel shows the prior acceptance rate using the set of priors given in Section 2 taking $\bm{\alpha} = (\alpha_1, \alpha_2, \alpha_3 + \alpha_4)$ where $\alpha_1 = 1, \alpha_2 = 1$ and increasing $\alpha_3 + \alpha_4$ from 2 to 100. Without instrumental variable assumptions priors on $\omega_{x,z}$ are $beta(\alpha_3, \alpha_4)$ distributions (becoming uniform when $\alpha_3 = \alpha_4 = 1$), thus increasing $\alpha_3$ and $\alpha_4$ while assuming they are equal results in beta distributions increasingly concentrated at 0.5. This causes less prior variability in $P(Y|R=0)$ and more prior samples with odds ratios close to 1 in our rejection sampler. 

Similarly, the proportion of missing data can affect posterior acceptance rates. We would expect as the proportion of missing data increases, our posterior acceptance rate to go down as we have less confidence in our instrumental variable assumption. This is true for models where we do not impose a strict instrumental variable assumption by setting $OR(Y,Z|X = 1)$ exactly, rather taking the OR to be in some interval around 1 (Figure 4b). However, where we do impose this exact assumption our rejection sampling algorithm changes slightly and the posterior acceptance rate increases with the proportion of missing data. This is because instead of sampling from independent Dirichlet and uniform/beta distributions and accepting samples that satisfy our assumptions, we sample from a uniform/beta distribution for $\omega_{x,1}$ and set $\omega_{x,0}$ such that the odds ratio is exactly 1. We do this because $OR(Y,Z|X = 1)=1$ is an area 0 set in the joint posterior parameter space, so we are forced to sample in a lower dimensional space to obtain any samples that satisfy this assumption. In priors with threshold or probabilistic assumptions on the instrumental variable, the acceptance rate is the proportion of samples from our saturated model that satisfy a given set of assumptions. However, using priors where we make the exact instrumental variable assumption the acceptance rate is the proportion where the assumptions are possible, and if possible we set parameters such that the assumptions are satisfied. Because the rejection sampling algorithm is the same in the prior and posterior, this may not be a problem, however as seen in Figure 4a careful consideration needs to be taken when setting priors. Given the results in Figure 4b, while possible to use an exact prior setting $OR(Y,Z|X = 1)$, in practice we would recommend using the softer prior instrumental variable assumptions given in the Sat-thresh or Sat-normal models, as was done in Section 5.

\clearpage

\section*{References}

\singlespacing

\begin{hangparas}{.5in}{1}

Balke, A., and Pearl, J. (1997), "Bounds on treatment effects from studies with imperfect compliance," \textit{Journal of the American Statistical Association}, 92, 1171-1176, doi: 10.1080/ \allowbreak 01621459.1997.10474074. 

Ban, K., and Kédagni, D. (2022), "Nonparametric bounds on treatment effects with imperfect instruments," \textit{The Econometrics Journal}, 25, 477-493, doi: 10.1093/ectj/utab033.

Daly‐Grafstein, D., and Gustafson, P. (2022), "Combining parametric and nonparametric models to estimate treatment effects in observational studies," \textit{Biometrics}, doi: 10.1111/biom.13776.

Daniels, M. J., and Hogan, J. W. (2008), Missing data in longitudinal studies: Strategies for Bayesian modeling and sensitivity analysis, CRC press.

Dwyer, R., Palepu, A., Williams, C., and Zhao, J. (2021), "Unconditional cash transfers reduce homelessness," \textit{psyarxiv}, doi: 10.31234/osf.io/ukngr.

Gustafson, P. (2015), Bayesian inference for partially identified models: Exploring the limits of limited data (Vol. 140), CRC Press.

Heckman, J. J. (1979), "Sample selection bias as a specification error," \textit{Econometrica: Journal of the econometric society}, 153-161.

Linero, A. R., and Daniels, M. J. (2015), "A flexible Bayesian approach to monotone missing data in longitudinal studies with nonignorable missingness with application to an acute schizophrenia clinical trial," \textit{Journal of the American Statistical Association}, 110, 45-55, doi: 10.1080/01621459.2014.969424.

Linero, A. R., and Daniels, M. J. (2018), "Bayesian approaches for missing not at random outcome data: The role of identifying restrictions," \textit{Statistical science: a review journal of the Institute of Mathematical Statistics}, 33, 198-213, doi: 10.1214/17-STS630.

Manski, C. F. (2005), "Partial identification with missing data: concepts and findings," \textit{International Journal of Approximate Reasoning}, 39, 151-165, doi:10.1016/j.ijar.2004.10.006. 

Marden, J. R., Wang, L., Tchetgen, E. J. T., Walter, S., Glymour, M. M., and Wirth, K. E. (2018), "Implementation of instrumental variable bounds for data missing not at random," \textit{Epidemiology}, 29, 364-368, doi: 10.1097/EDE.0000000000000811.

Mattei, A., Mealli, F., and Pacini, B. (2014). "Identification of causal effects in the presence of nonignorable missing outcome values," \textit{Biometrics}, 70, 278-288, doi: 10.1111/biom.12136

McGovern, M. E., Bärnighausen, T., Marra, G., and Radice, R. (2015), "On the assumption of bivariate normality in selection models: a copula approach applied to estimating HIV prevalence," \textit{Epidemiology}, 26, 229-37, doi: 10.1097/EDE.0000000000000218.

Little, R. J. (1993), "Pattern-mixture models for multivariate incomplete data," \textit{Journal of the American Statistical Association}, 88, 125-134, doi: 10.1080/01621459.1993.10594302.

Puhani, P. (2000), "The Heckman correction for sample selection and its critique," \textit{Journal of economic surveys}, 14, 53-68, doi: 10.1111/1467-6419.00104.

Rescorla, M. (2015), "Some epistemological ramifications of the Borel–Kolmogorov paradox," \textit{Synthese}, 192, 735-767, doi: 10.1007/s11229-014-0586-z.

Rubin, D. B. (1976), "Inference and missing data," \textit{Biometrika}, 63, 581-592, doi: 10.1093/biomet/63.3.581.

Swanson, S. A., Hernán, M. A., Miller, M., Robins, J. M., and Richardson, T. S. (2018), "Partial identification of the average treatment effect using instrumental variables: review of methods for binary instruments, treatments, and outcomes," \textit{Journal of the American Statistical Association}, 113, 933-947, doi: 10.1080/01621459.2018.1434530.

Van de Ven, W. P., and Van Praag, B. M. (1981), "The demand for deductibles in private health insurance: A probit model with sample selection," \textit{Journal of econometrics}, 17, 229-252, doi: 10.1016/0304-4076(81)90028-2.

Van Hasselt, M. (2011), "Bayesian inference in a sample selection model," \textit{Journal of Econometrics}, 165, 221-232, doi: 10.1016/j.jeconom.2011.08.003.

\end{hangparas}

\clearpage

\doublespacing

\setcounter{section}{0} 
\setcounter{equation}{0} 
\renewcommand\thesection{\Alph{section}} 

\section{Appendix}

\renewcommand{\theequation}{\thesection.\arabic{equation}}

\subsection{Missing data restrictions for imperfect instruments}

If we define our imperfect instrument as a restriction on $OR(Y,Z|X)$ to be within some threshold $(t_l, t_h)$ around 1, we can write the joint prior as

\begin{align}
f(q_{0,0},q_{0,1},q_{1,0},q_{1,1},\omega_{0,0},\omega_{0,1},\omega_{1,0},\omega_{1,1}) \propto &f(q_{0,0} ; \bm{\alpha}_{0,0}) f(q_{0,1} ; \bm{\alpha}_{0,1})f(q_{1,0} ; \bm{\alpha}_{1,0})f(q_{1,1} ; \bm{\alpha}_{1,1}) \nonumber \\
&I_{(0,1)}(\omega_{0,0})I_{(0,1)}(\omega_{0,1})I_{(0,1)}(\omega_{1,0})I_{(0,1)}(\omega_{1,1}) \\ 
&I\{OR(Y,Z|X=1) \in (t_l, t_h)\} I\{OR(Y,Z|X=0) \in (t_l, t_h)\}  \nonumber.
\end{align}

The constraint in (3), equivalently written on the odds ratio scale rather than the risk difference scale, is now

\begin{equation}
 t_l < \frac{\frac{q_{0,1,1,1} + w_{0,1}q_{0,1,0,\boldsymbol{\cdot}}}{1 - (q_{0,1,1,1} + w_{0,1}q_{0,1,0,\boldsymbol{\cdot}})}}{\frac{q_{0,0,1,1} + w_{0,0}q_{0,0,0,\boldsymbol{\cdot}}}{1 - (q_{0,0,1,1} + w_{0,0}q_{0,0,0,\boldsymbol{\cdot}})}} < t_h.
\end{equation}

Similar to the Sat-OR-1 model, we can derive nonparametric bounds for $\Psi$. Consider the assumption (A.2) when $\omega_{0,0} = \omega_{0,1} = 0$. If

$$ t_l < \frac{\frac{q_{0,1,1,1}}{1 - q_{0,1,1,1}}}{\frac{q_{0,0,1,1}}{1 - q_{0,0,1,1}}} < t_h $$,

then our instrumental variable assumption is satisfied and these are our lower bounds for $\omega_{0,0}$ and $\omega_{0,1}$. In the case where

\begin{equation}
\frac{\frac{q_{0,1,1,1}}{1 - q_{0,1,1,1}}}{\frac{q_{0,0,1,1}}{1 - q_{0,0,1,1}}} < t_l 
\end{equation}

we need to increase the lower bound for $\omega_{0,1}$ to get within our interval. Setting

$$ \frac{\frac{q_{0,1,1,1} + q_{0,1,0,\boldsymbol{\cdot}} \omega_{0,1}}{1 - (q_{0,1,1,1} + q_{0,1,0,\boldsymbol{\cdot}} \omega_{0,1})}}{\frac{q_{0,0,1,1}}{1 - q_{0,0,1,1}}} = t_l $$

and solving for $\omega_{0,1}$ we get

\begin{equation}
\omega_{0,1} =  \frac{q_{0,0,1,1} t_l - q_{0,1,1,1} q_{0,0,1,1} t_l - q_{0,1,1,1} + q_{0,0,1,1} q_{0,1,1,1}}{{q_{0,1,0,\boldsymbol{\cdot}}} (1 - q_{0,0,1,1} + q_{0,0,1,1} t_l)}.
\end{equation}

Similarly, when 

\begin{equation}
\frac{\frac{q_{0,1,1,1}}{1 - q_{0,1,1,1}}}{\frac{q_{0,0,1,1}}{1 - q_{0,0,1,1}}} > t_h
\end{equation}

we set

$$ \frac{\frac{q_{0,1,1,1}}{1 - q_{0,1,1,1}}}{\frac{q_{0,0,1,1} + q_{0,0,0,\boldsymbol{\cdot}} \omega_{0,0}}{1 -(q_{0,0,1,1} + q_{0,0,0,\boldsymbol{\cdot}} \omega_{0,0} )}} = t_h $$.

Solving for $\omega_{0,0}$ we get

\begin{equation}
\omega_{0,0} = \frac{q_{0,0,1,1} t_h - q_{0,1,1,1} q_{0,0,1,1} t_h - q_{0,1,1,1} + q_{0,0,1,1} q_{0,1,1,1}}{{q_{0,0,0,\boldsymbol{\cdot}}} (q_{0,1,1,1} t_h - q_{0,1,1,1} - t_h)}.
\end{equation}

Note when (A.3) is greater than $t_l$, then (A.4) is less than 0. Similarly when (A.5) is less than $t_h$ then (A.6) is less than 0. Therefore our lower bounds for $f(\omega_{0,1}|{q_{x,z}})$ and $f(\omega_{0,0}|{q_{x,z}})$ are

\begin{align*}
\omega_{0,1} = &max \left (0,\frac{q_{0,0,1,1} t_l - q_{0,1,1,1} q_{0,0,1,1} t_l - q_{0,1,1,1} + q_{0,0,1,1} q_{0,1,1,1}}{{q_{0,1,0,\boldsymbol{\cdot}}} (1 - q_{0,0,1,1} + q_{0,0,1,1} t_l)} \right) \\
\omega_{0,0} = &max \left (0, \frac{q_{0,0,1,1} t_h - q_{0,1,1,1} q_{0,0,1,1} t_h - q_{0,1,1,1} + q_{0,0,1,1} q_{0,1,1,1}}{{q_{0,0,0,\boldsymbol{\cdot}}} (q_{0,1,1,1} t_h - q_{0,1,1,1} - t_h)} \right)
\end{align*}

Similar reasoning gives upper bounds for $f(\omega_{0,1}|{q_{x,z}})$ and $f(\omega_{0,0}|{q_{x,z}})$ of

\begin{align*}
\omega_{0,1} = &min \left (1, \frac{q_{0,0,1,1} t_h - q_{0,1,1,1} q_{0,0,1,1} t_h + {q_{0,1,0,\boldsymbol{\cdot}}} t_h - {q_{0,0,0,\boldsymbol{\cdot}}} q_{0,1,1,1} t_h - q_{0,1,1,1} + q_{0,0,1,1} q_{0,1,1,1} + {q_{0,0,0,\boldsymbol{\cdot}}}  q_{0,1,1,1}}{{q_{0,1,0,\boldsymbol{\cdot}}} (1 - q_{0,0,1,1} - {q_{0,0,0,\boldsymbol{\cdot}}} + q_{0,0,1,1} t_h + {q_{0,0,0,\boldsymbol{\cdot}}} t_h)}\right ) \\
\omega_{0,0} = &min \left (1, \frac{q_{0,0,1,1} t_l - q_{0,1,1,1} q_{0,0,1,1} t_l - {q_{0,1,0,\boldsymbol{\cdot}}} q_{0,0,1,1} t_l - q_{0,1,1,1} + q_{0,0,1,1} q_{0,1,1,1} - {q_{0,1,0,\boldsymbol{\cdot}}} + {q_{0,1,0,\boldsymbol{\cdot}}} q_{0,0,1,1}}{{q_{0,0,0,\boldsymbol{\cdot}}} ({q_{0,1,0,\boldsymbol{\cdot}}} t_l + q_{0,1,1,1} t_l - t_l - {q_{0,1,0,\boldsymbol{\cdot}}} - q_{0,1,1,1})}\right ).
\end{align*}

Bounds are similar for $\omega_{1,0}, \omega_{1,1}$ and can be plugged into (5) to get the resulting nonparametric bounds on $\Psi$. 

We can also characterize our prior assumptions about how close $Z$ is to a true instrument by inducing a smooth but concentrated probability distribution over $OR(Y,Z|X=1)$ and $OR(Y,Z|X=0)$. Here we give the example of multiplying the prior in (1) by normal density functions over $log(OR(Y,Z|X))$ such that the standard deviation of these normals $\sigma$ determines how close we think $Z$ to a true instrument. This gives a joint prior of

\begin{align}
f(q_{0,0},q_{0,1},q_{1,0},q_{1,1},\omega_{0,0},\omega_{0,1},\omega_{1,0},\omega_{1,1}) \propto &f(q_{0,0} ; \bm{\alpha}_{0,0}) f(q_{0,1} ; \bm{\alpha}_{0,1})f(q_{1,0} ; \bm{\alpha}_{1,0})f(q_{1,1} ; \bm{\alpha}_{1,1}) \nonumber \\
&I_{(0,1)}(\omega_{0,0})I_{(0,1)}(\omega_{0,1})I_{(0,1)}(\omega_{1,0})I_{(0,1)}(\omega_{1,1}) \\ 
&\phi(log(OR(Y,Z|X=1)); 0, \sigma) \nonumber \\
&\phi(log(OR(Y,Z|X=0)); 0, \sigma). \nonumber
\end{align}

When using the prior (A.7) the nonparametric bounds are the same as when no assumptions are made, since any odds ratio is possible. However, the restrictions given when we restrict the odds ratio to an interval $(t_l, t_h)$ can give us an idea of the shape of the priors on $\omega_{x,z}$ in this case. For example, if we assume $\sigma = log(t_h)/2$ and we have $t_l = 1/t_h$, then we can expect priors on $\omega_{x,z}$ where approximately $95\%$ of the mass of the distributions are within the bounds given for $\omega_{x,z}$ in the Sat-OR-thresh model above.

\subsection{Prior for the Sat-OR-normal-beta model}

Taking our transformed missing data biases $h(\omega_{x,z}|q_{x,z})$ as in Section 2.2 we get a joint prior 

\begin{align}
f(q_{0,0},q_{0,1},q_{1,0},q_{1,1},\omega_{0,0},\omega_{0,1},\omega_{1,0},\omega_{1,1}) \propto &f(q_{0,0} ; \bm{\alpha}_{0,0}) f(q_{0,1} ; \bm{\alpha}_{0,1})f(q_{1,0} ; \bm{\alpha}_{1,0})f(q_{1,1} ; \bm{\alpha}_{1,1}) \nonumber \\
&I_{(0,1)}(\omega_{0,0})I_{(0,1)}(\omega_{0,1})I_{(0,1)}(\omega_{1,0})I_{(0,1)}(\omega_{1,1}) \\ 
&\phi(log(OR(Y,Z|X=1)): 0, \sigma) \nonumber \\
&\phi(log(OR(Y,Z|X=0)): 0, \sigma). \nonumber \\
&f_h(\omega_{0,0}|q_{0,0})f_h(\omega_{0,1}|q_{0,1}) \nonumber \\
&f_h(\omega_{1,0}|q_{1,0})f_h(\omega_{1,1}|q_{1,1}), \nonumber 
\end{align}

where $f_h$ is the probability density of a $beta(a,b)$ distribution. 

\subsection{Gibbs sampler for the Heckman model}

We provide the full algorithm used for our Gibbs sampler in Section 4. This algorithm is a modification of the algorithm given in Van Hasselt (2011) for discrete outcome data. 

For convenience, we use a slight change of notation from Section 1.2. Let $R_i^* = C_{i1}^T \bm{\gamma} + \nu_i$ with $C_{i1}^T = (1, X_i, Z_i)$, and $Y_i^* = C_{i2}^T \bm{\beta} + \epsilon_i$ with $C_{i2}^T = (1, X_i)$. Let $\theta = (\bm{\gamma}, \bm{\beta}, \rho)$, $N_1 = \{i: R_i = 1 \}$, $N_0 = \{i: R_i = 0\}$. We assume priors of $\bm{\gamma} \sim MVN(b_1, B_1)$,  $\bm{\beta} \sim MVN(b_2, B_2)$, $\rho \sim Unif(-1, 1)$.

\begin{enumerate}
  \item Initialize $r^{*(0)}, y^{*(0)}, \theta^{(0)}$
  \item Sample from 

\begin{align*}
[r_i^*|y_i^*, r_i, \theta] \sim  &\mathcal{TN}_{(-\infty, 0)} (C_{i1}^T \bm{\gamma}, 1), \hspace{0.25cm} i \in N_0 \\
&\mathcal{TN}_{(0, \infty)} (C_{i1}^T \bm{\gamma} + \rho(y_i^* - C_{i2}^T\bm{\beta}), 1 - \rho^2), \hspace{0.25cm} i \in N_1
\end{align*}

  \item  Sample from 

\begin{align*}
[y_i^*|r_i^*, y_i, \theta] \sim  &\mathcal{TN}_{(-\infty, 0)} (C_{i2}^T \bm{\beta} + \rho(r_i^* - C_{i1}^T \bm{\gamma}) , 1 - \rho^2), \hspace{0.25cm} y_i = 0 \\
&\mathcal{TN}_{(0, \infty)} (C_{i2}^T \bm{\beta} + \rho(r_i^* - C_{i1}^T \bm{\gamma}) , 1 - \rho^2) \hspace{0.25cm}, y_i = 1
\end{align*}

  \item  Sample from 
$$[\bm{\gamma}|r_i^*, y_i^* \bm{\beta}, \rho] \sim N(\bar{b}_1, \bar{B}_1)$$

where

\begin{align*}
\bar{B}_1 & = \left( B_1^{-1} + \sum_{i \in N_0} C_{i1}C_{i1}^T + (1-\rho^2)^{-1}\sum_{i \in N_1} C_{i1}C_{i1}^T \right)^{-1} \\
\bar{b}_1 & = \bar{B}_1\left( B_1^{-1}b_1 + \sum_{i \in N_0} C_{i1}r_i^* + (1-\rho^2)^{-1}\sum_{i \in N_1} C_{i1}(r_i^* - \rho(y_i^* - C_{i2}^T\bm{\beta})) \right)^{-1}
\end{align*}

  \item  Sample from 
$$[\bm{\beta}|r_i^*, y_i^* \bm{\gamma}, \rho] \sim N(\bar{b}_2, \bar{B}_2)$$

where

\begin{align*}
\bar{B}_2 & = \left( B_2^{-1} + (1-\rho^2)^{-1}\sum_{i \in N_1} C_{i2}C_{i2}^T \right)^{-1} \\
\bar{b}_2 & = \bar{B}_2\left( B_2^{-1}b_2 + (1-\rho^2)^{-1}\sum_{i \in N_1} C_{i2}(y_i^* - \rho(r_i^* - C_{i1}^T\bm{\gamma})) \right)^{-1}
\end{align*}

  \item  Sample from $[\rho|r_i^*, y_i^* \bm{\beta}, \bm{\gamma}]$ using a Metropolis-Hastings step. We use a proposal of $\rho ^\prime|\rho \sim N(\rho, 0.05)$  and an acceptance probability of 

$$ \alpha(\rho,\rho^\prime) = min\left( \frac{f(y^*, r^*| \bm{\gamma}, \bm{\beta}, \rho^\prime)I(\rho^\prime \in (-1,1))}{f(y^*, r^*| \bm{\gamma}, \bm{\beta}, \rho)I(\rho \in (-1,1))}, 1 \right)$$

\end{enumerate}

\subsection{Full results for simulation 2}

\begin{table}[!t]
    \scriptsize
    \centering
    \caption{Mean acceptance rate, (geometric) mean Bayes factor, proportion of posteriors computed, coverage, and mean 90\% credible interval width over 200 runs from simulation 2. Note coverage and credible intervals are calculated on the subset of runs where the posterior is computed successfully. The "Heckman" model assumes a Bayesian Heckman selection model, and the "Oracle" model fits a saturated model without missing data. Data are simulated either using a Heckman DGP or a saturated DGP with one or both of the instrumental variable and positive data bias assumptions violated. All Bayes factors compare the posterior odds of the "Sat" model to the model of interest.}
    \vspace{0.3cm}
    \label{tab:mytable}
    \renewcommand{\arraystretch}{1.03} 
    \begin{tabular}{|c|c|c|c|c||c|c|}
        \hline
        \multirow{2}{*}{\textbf{Simulation}} & \multirow{2}{*}{\textbf{Model}} & \multicolumn{5}{c|}{\textbf{Results}} \\
        \cline{3-7}
        & & \textbf{AR} & \textbf{BF} & \textbf{Computed} & \textbf{Coverage} & \textbf{Interval width} \\
        \hline
        \multirow{5}{*} & Sat & &  &  & 0.98 & 0.21   \\
         & Sat-OR-1 & 1.8e-03& 798 & 0.23 & 0.89 & 0.14  \\
        Saturated DGP & Sat-OR-thresh & 4.6e-03& 63.9 & 0.41 & 0.88 & 0.15  \\
        $P(R=0) = 0.2$ & Sat-OR-normal & 0.017 & 5.9 & 0.95 & 0.92 & 0.18  \\
        $OR \neq 1$, pos bias  & Sat-OR-1-pos & 2.5e-07 & 5750 & 0 & NA & NA  \\
        & Beta bias & 8.8e-04 & 14.7  & 0.27  & 0.91 & 0.18   \\
        & Heckman & &  &  & 0.73 & 0.10   \\
        & Oracle & &  &  & 0.88 & 0.10  \\
        \hline
        \multirow{5}{*} & Sat & &  &  & 1 & 0.39   \\
         & Sat-OR-1 & 0.24 & 0.72 & 1 & 0.91 & 0.31 \\
        Saturated DGP & Sat-OR-thresh & 0.055 & 0.66 & 1 & 0.94 & 0.33  \\
        $P(R=0) = 0.4$ & Sat-OR-normal &  0.081 & 0.65 & 1 & 0.95 & 0.35  \\
        $OR \neq 1$, pos bias  & Sat-OR-1-pos & 7.8e-03 & 93.7 & 0.26 & 0.82 & 0.36  \\
        & Sat-OR-normal-beta & 6.1e-03 & 1.0  & 1  & 1 & 0.32   \\
        & Heckman & &  &  & 0.41 & 0.11   \\
        & Oracle & &  &  & 0.87 & 0.10  \\
        \hline
        \multirow{5}{*} & Sat & &  &  & 0.98 & 0.24   \\
         & Sat-OR-1 & 4.5e-03 & 1060 & 0.24 & 0.67 & 0.14  \\
        Saturated DGP & Sat-OR-thresh & 5.5e-03 & 28.0 & 0.53 & 0.71 & 0.15  \\
        $P(R=0) = 0.2$ & Sat-OR-normal & 0.019 & 4.33 & 1 & 0.90 & 0.18  \\
        $OR \neq 1$, neg bias  & Sat-OR-1-pos & 2.2e-06 & 261 & 0 & NA & NA  \\
        & Sat-OR-normal-beta & 8.1e-04 & 17.7  & 0.23 & 1 & 0.19   \\
        & Heckman & &  &  & 0.68 & 0.11   \\
        & Oracle & &  &  & 0.89 & 0.10  \\
        \hline
        \multirow{5}{*} & Sat & &  &  & 1 & 0.39   \\
         & Sat-OR-1 & 0.27 & 1 & 1 & 0.93 & 0.32  \\
        Saturated DGP & Sat-OR-thresh & 0.061 & 0.64 & 1 & 0.98 & 0.34  \\
        $P(R=0) = 0.4$ & Sat-OR-normal & 0.090 & 0.61 & 1 & 0.98 & 0.35  \\
        $OR \neq 1$, neg bias  & Sat-OR-1-pos & 4.0e-03 & 92.6 & 0.28 & 0.91 & 0.28  \\
        & Sat-OR-normal-beta & 3.2e-03 & 2.14 & 0.80  & 0.97 & 0.31   \\
        & Heckman & &  &  & 0.38 & 0.11   \\
        & Oracle & &  &  & 0.90 & 0.10  \\
        \hline
        \multirow{5}{*} & Sat & &  &  & 0.95 & 0.21   \\
         & Sat-OR-1 & 0.20 & 1.43 & 0.97 & 0.95 & 0.19  \\
        Saturated DGP & Sat-OR-thresh & 0.17 & 0.29 & 1 & 0.95 & 0.20  \\
        $P(R=0) = 0.2$ & Sat-OR-normal & 0.23 & 0.27 & 1 & 0.97 & 0.21  \\
        $OR = 1$, neg bias  & Sat-OR-1-pos & 0.028 & 19.6 & 0.43 & 0.86 & 0.18  \\
        & Sat-OR-normal-beta & 0.022 & 0.42  & 0.94  & 0.97 & 0.19   \\
        & Heckman & &  &  & 0.58 & 0.11   \\
        & Oracle & &  &  & 0.92 & 0.10  \\
        \hline
        \multirow{5}{*} & Sat & &  &  & 0.87 & 0.38   \\
         & Sat-OR-1 & 0.30 & 0.85 & 1 & 0.88 & 0.34  \\
        Saturated DGP & Sat-OR-thresh & 0.069 & 0.71 & 1 & 0.91 & 0.34  \\
        $P(R=0) = 0.4$ & Sat-OR-normal & 0.096 & 0.67 & 1 & 0.92 & 0.35  \\
        $OR = 1$, neg bias  & Sat-OR-1-pos & 0.075 & 2.51 & 0.46 & 0.59 & 0.35  \\
        & Sat-OR-normal-beta & 9.1e-03 & 1.84  & 0.72  & 0.79 & 0.32   \\
        & Heckman & &  &  & 0.25 & 0.12   \\
        & Oracle & &  &  & 0.90 & 0.10  \\
        \hline
    \end{tabular}
\end{table}

\begin{table*}[!t]
     \scriptsize
     \centering
     \caption*{}
    \vspace{0.3cm}
    \label{tab:mytable}
    \renewcommand{\arraystretch}{1.03} 
    \begin{tabular}{|c|c|c|c|c||c|c|}
        \hline
        \multirow{2}{*}{\textbf{Simulation}} & \multirow{2}{*}{\textbf{Model}} & \multicolumn{5}{c|}{\textbf{Results}} \\
        \cline{3-7}
        & & \textbf{AR} & \textbf{BF} & \textbf{Computed} & \textbf{Coverage} & \textbf{Interval width} \\
        \hline
        \multirow{5}{*} & Sat & &  &  & 1 & 0.24   \\
         & Sat-OR-1 & 9.0e-06 & 19221 & 0 & NA & NA  \\
        Heckman DGP & Sat-OR-thresh & 2.0e-04& 521 & 0.025 & 1 & 0.14  \\
        $P(R=0) = 0.2$ & Sat-OR-normal & 5.3e-03 & 12.3 & 0.95 & 0.71 & 0.16  \\
        $OR \neq 1$, pos bias  & Sat-OR-1-pos & 0 & NA & 0 & NA & NA  \\
        & Sat-OR-normal-beta & 1.5e-04 & 53.1  & 0  & NA & NA   \\
        & Heckman & &  &  & 0.24 & 0.10   \\
        & Oracle & &  &  & 0.94 & 0.10  \\
        \hline
        \multirow{5}{*} & Sat & &  &  & 1 & 0.39   \\
         & Sat-OR-1 & 0.24 & 0.76 & 1 & 0.96 & 0.31 \\
        Heckman DGP & Sat-OR-thresh & 0.041 & 0.90 & 1 & 1 & 0.32  \\
        $P(R=0) = 0.4$ & Sat-OR-normal &  0.063 & 0.84 & 1 & 1 & 0.33  \\
        $OR \neq 1$, pos bias  & Sat-OR-1-pos & 2.2e-05 & 2234 & 0 & NA & NA  \\
        & Sat-OR-normal-beta & 2.8e-03 & 2.08  & 0.54  & 1 & 0.30   \\
        & Heckman & &  &  & 0.38 & 0.10   \\
        & Oracle & &  &  & 0.88 & 0.10  \\
        \hline
        \multirow{5}{*} & Sat & &  &  & 1 & 0.23   \\
         & Sat-OR-1 & 1.3e-03 & 626 & 0.25 & 0 & 0.13  \\
        Heckman DGP & Sat-OR-thresh & 3.0e-03 & 37.4 & 0.61 & 0.08 & 0.14  \\
        $P(R=0) = 0.2$ & Sat-OR-normal & 0.015 & 4.49 & 1 & 0.41 & 0.16  \\
        $OR \neq 1$, neg bias  & Sat-OR-1-pos & 0 & NA & 0 & NA & NA  \\
        & Sat-OR-normal-beta & 2.1e-04 & 44.7  & 0.035  & 0 & 0.16   \\
        & Heckman & &  &  & 0.06 & 0.09   \\
        & Oracle & &  &  & 0.91 & 0.10  \\
        \hline
        \multirow{5}{*} & Sat & &  &  & 1 & 0.39   \\
         & Sat-OR-1 & 0.36 & 0.46 & 1 & 1 & 0.33  \\
        Heckman DGP & Sat-OR-thresh & 0.05 & 0.67 & 1 & 1 & 0.33  \\
        $P(R=0) = 0.4$ & Sat-OR-normal & 0.077 & 0.67 & 1 & 1 & 0.35  \\
        $OR \neq 1$, neg bias  & Sat-OR-1-pos & 1.5e-06 & 1412 & 0 & NA & NA  \\
        & Sat-OR-normal-beta & 7.6e-04 & 7.4 & 0.15  & 1 & 0.28   \\
        & Heckman & &  &  & 0.83 & 0.08   \\
        & Oracle & &  &  & 0.93 & 0.09  \\
        \hline
        \multirow{5}{*} & Sat & &  &  & 0.95 & 0.24   \\
         & Sat-OR-1 & 0.85 & 0.19 & 1 & 0.95 & 0.20  \\
        Heckman DGP & Sat-OR-thresh & 0.32 & 0.11 & 1 & 0.97 & 0.21  \\
        $P(R=0) = 0.2$ & Sat-OR-normal & 0.37 & 0.14 & 1 & 0.98 & 0.22  \\
        $OR = 1$, neg bias  & Sat-OR-1-pos & 0.053 & 0.68 & 0.98 & 0.92 & 0.16  \\
        & Sat-OR-normal-beta & 0.031 & 0.17  & 1 & 0.93 & 0.21   \\
        & Heckman & &  &  & 0.86 & 0.11   \\
        & Oracle & &  &  & 0.90 & 0.10  \\
        \hline
        \multirow{5}{*} & Sat & &  &  & 1 & 0.40   \\
         & Sat-OR-1 & 0.95 & 0.17 & 1 & 1 & 0.45  \\
        Heckman DGP & Sat-OR-thresh & 0.14 & 0.24 & 1 & 1 & 0.43  \\
        $P(R=0) = 0.4$ & Sat-OR-normal & 0.19 & 0.26 & 1 & 1 & 0.43  \\
        $OR = 1$, neg bias  & Sat-OR-1-pos & 0.13 & 0.18 & 1 & 0.99 & 0.26  \\
        & Sat-OR-normal-beta & 0.014 & 0.36  & 1  & 1 & 0.34   \\
        & Heckman & &  &  & 0.83 & 0.13   \\
        & Oracle & &  &  & 0.87 & 0.10  \\
        \hline
    \end{tabular}
\end{table*}

\end{document}